# Application of Spectroscopic Ellipsometry and Mueller Ellipsometry to Optical Characterization

E. Garcia-Caurel[1], A. De Martino[1], J-P. Gaston[2], L.Yan[3]
[1]LPICM, CNRS-Ecole Polytechnique, Palaiseau, France
[2]HORIBA Scientific, Chilly-Mazarin, France
[3]HORIBA Scientific, Edison, NJ, USA

This article aims to provide a brief overview of both established and novel ellipsometry techniques, as well as their applications. Ellipsometry is an indirect optical technique in that information about the physical properties of a sample is obtained through modeling analysis. Standard ellipsometry is typically used to characterize optically isotropic bulk and/or layered materials. More advanced techniques like Mueller ellipsometry, also known as polarimetry in literature, are necessary for the complete and accurate characterization of anisotropic and/or depolarizing samples which occur in many instances, both in research and "real life" activities. In this article we cover three main areas of subject: basic theory of polarization, standard ellipsometry and Mueller ellipsometry. Section I is devoted to a short and pedagogical introduction of the formalisms used to describe light polarization. The following section is devoted to standard ellipsometry. The focus is on the experimental aspects, including both pros and cons of commercially available instruments. Section III is devoted to recent advances in Mueller ellipsometry. Applications examples are provided in sections II and III to illustrate how each technique works.

## INTRODUCTION

The use of polarized light to characterize the optical properties of materials, either in bulk or thin film format, has enjoyed great success over the past decades. The different methods of generating and analyzing the polarization properties of light is traditionally called Ellipsometry. The particularity of spectroscopic ellipsometry is that it measures two independent values at each wavelength, allowing the technique to provide more information than other available techniques, such as conventional reflectometry. This makes spectroscopic ellipsometry a highly accurate thin film measurement tool. The technique finds its roots in the pioneering work of Paul Drude in the 19th century, when he used a polarized light in reflection configuration to study the optical properties and thickness of very thin metallic films. Since then, thousands of studies and industrial applications have emerged, which are based either on ellipsometry, or profit directly from its sensitivity. In spite of its sensitivity, ellipsometric measurements require that the light beam remains completely polarized during the measurement. A beam of light is said to be polarized when the relative phase between the different components of the electromagnetic field (associated with the light beam) remain related to each other in a deterministic and predictable way. If for some reason the phase relation is perturbed, light will become partially polarized, so the ellipsometric measurements lose their physical meaning. To take into account the partially polarized light, it is necessary to use the more general technique called Mueller Ellipsometry or Polarimetry. We prefer the term 'Mueller Ellipsometry' because it shows the close relation with standard ellipsometry. Several excellent monographs[1-4] have been published covering different aspects of both standard and Mueller Ellipsometry, including the theory of polarization, the optical response of solids, instrumentation and innovative applications. The goal of this article is not to summarize the information contained in these monographs, but rather to give an overview of the state-of-the-art technologies developed in the framework of the company HORIBA Scientific, in collaboration with several research laboratories.

The article is organized in three sections. The first section gives an overview of the optical formalisms used to describe the polarization state of light. The second and third sections are devoted to standard and Mueller Ellipsometry, respectively. We will describe the prominent features of some instruments, in particular a phase-modulated spectroscopic ellipsometer, a spectroscopic Mueller ellipsometer and an angle-resolved imaging Mueller ellipsometer. We will also provide some examples to illustrate the use and performance of each instrument.

**Polarization of light**
In this section, we briefly review the most widely used theoretical descriptions of the light polarization properties, namely the Jones formalism for fully polarized light and the Stokes-Mueller formalism, which is the most general representation, and can adequately account for any polarization states. The polarimetric properties of any sample are then defined from the changes this sample introduces in the polarization state of light. In turn, these properties may be used for various purposes, from the very well established (such as material and thin film characterizations), to more advanced applications, such as remote sensing and/or medical diagnosis.

As described in any textbooks on electromagnetism[5], when a light ray propagates (through an isotropic medium) along the $z$ direction, the electric field vector **E** is confined to vibrate in an $xy$ plane perpendicular to $z$, as illustrated in fig. 1.

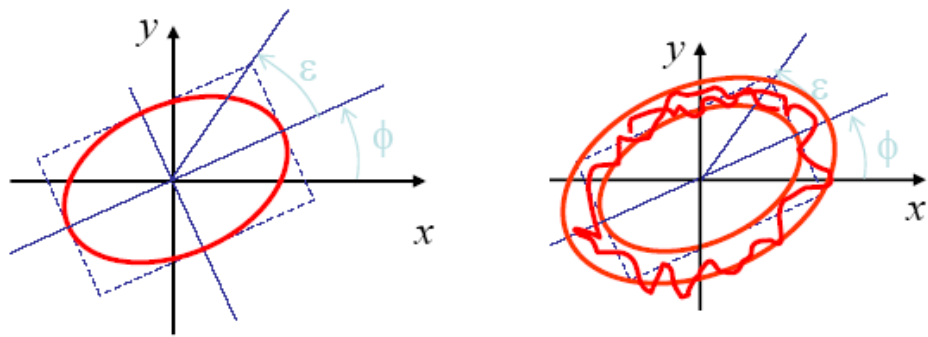

Figure 1. Examples of electric field trajectories in the plane perpendicular to the propagation direction for fully polarized (*left*) or partially polarized (*right*) light waves.

For fully polarized light, the electric field **E** describes an ellipse, characterized by its ellipticity ($\varepsilon$) and the azimuth ($\phi$) of its major axis. The particular cases of linear and circular polarizations correspond to $\varepsilon = 0$ and $\phi = 45°$, respectively. In contrast, partially polarized states correspond to more disordered motions of the electric field, which can only be described statistically from cross-correlation functions as discussed below.

**The Jones Formalism**
As mentioned above, the Jones formalism is well adapted to the description of fully polarized states. Any elliptical motion of **E** can be decomposed along the $x$ and $y$ axes, with real amplitudes $A_i$ and phases $\phi_i$ ($i = x, y$), which can be lumped into complex numbers $E_i$ and form the Jones vector given by:

$$\begin{pmatrix} E_x \\ E_y \end{pmatrix} = \begin{pmatrix} A_x e^{i\varphi_x} \\ A_y e^{i\varphi_y} \end{pmatrix} \quad (1)$$

The Jones vector also contains an overall phase factor which may be important in some cases; e.g., when the polarized beam under study interferes with another beam. However, as long as only single-beam ellipsometry is concerned, this overall phase can be removed by setting $\phi_x = 0$. In the absence of depolarization, the interaction with a sample transforms the Jones vector of the incident beam into another Jones vector through a linear transformation:

$$\begin{pmatrix} E_x^{out} \\ E_y^{out} \end{pmatrix} = \begin{pmatrix} J_{xx} & J_{xy} \\ J_{yx} & J_{yy} \end{pmatrix} \begin{pmatrix} E_x^{in} \\ E_y^{in} \end{pmatrix} \quad (2)$$

where the $J_{ij}$ are the elements of the Jones matrix. In a similar way as for Jones vectors, if one is interested only in the polarimetric properties of the sample and not its overall optical path (or phase shift), one element can be taken as real; the Jones matrix which depends on seven real parameters further reduces to six if the overall amplitude transmission (or reflectivity) is also neglected.

For plane and isotropic samples, the Jones matrix in Eq. 2 takes on a special simple form: diagonal. It turns out that in practice the majority of substrates and thin films produced in research and industrial laboratories are isotropic, which makes the study by ellipsometry relatively straightforward. For this type of samples the two non-vanishing Jones matrix elements can be written in terms of the two Fresnel complex coefficients for the polarization, p, parallel and, s, perpendicular to the plane of incidence.

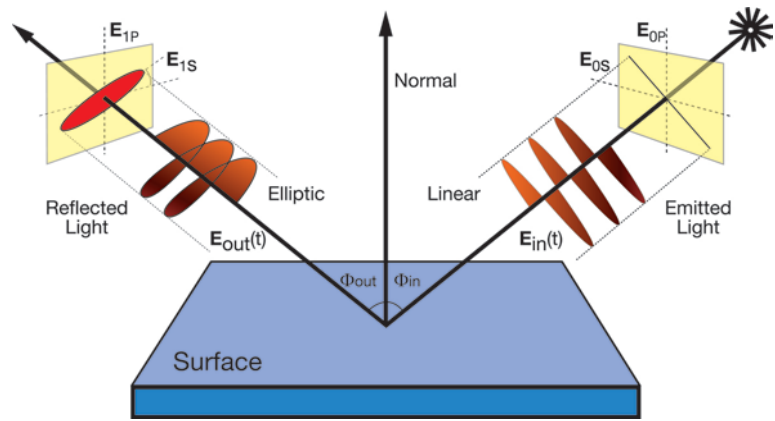

Figure 2. Schematic view of an ellipsometric measurement in reflection configuration. The polarized beam is incident on the sample from the right. After reflection, the polarization state of the beam is changed and light propagates to the left.

$$\begin{pmatrix} E_x^{out} \\ E_y^{out} \end{pmatrix} = \begin{pmatrix} J_{xx} & 0 \\ 0 & J_{yy} \end{pmatrix} \begin{pmatrix} E_x^{in} \\ E_y^{in} \end{pmatrix} = \begin{pmatrix} r_p & 0 \\ 0 & r_s \end{pmatrix} \begin{pmatrix} E_x^{in} \\ E_y^{in} \end{pmatrix}$$

In the particular case of samples consisting of a semi-infinite substrate of complex index $N_1=n_1+ik_1$, the complex Fresnel coefficients can be written as follows:

$$r_p = \frac{E_p^r}{E_p^i} = \frac{\tilde{N}_1\cos\phi_0 - \tilde{N}_0\cos\phi_1}{\tilde{N}_1\cos\phi_0 + \tilde{N}_0\cos\phi_1} = |r_p|e^{i\delta_p} \quad \text{and} \quad r_s = \frac{E_s^r}{E_s^i} = \frac{\tilde{N}_1\cos\phi_0 - \tilde{N}_1\cos\phi_1}{\tilde{N}_0\cos\phi_0 + \tilde{N}_1\cos\phi_1} = |r_s|e^{i\delta_s}$$

Where the complex index $N_0=n_0+ik_0$ represent the refractive index n and the absorption coefficient of the medium surrounding the sample (the air in general). The ellipsometric angles $\Psi$ and $\Delta$ are defined from the ratio of the complex Fresnel coefficients as:

$$\rho = \frac{r_p}{r_s} = \tan\Psi e^{i\Delta} \text{ where: } \tan\Psi = \frac{|r_p|}{|r_s|} \text{ et } \Delta = \delta_p - \delta_s$$

Thus, $\tan\Psi$ corresponds to the amplitude ratio upon reflection, and $\Delta$ is the difference in phase shift. Similar expression can be obtained for measurements performed in transmission by substitution of the Fresnel coefficients in reflection by the corresponding ones in transmission.

**The Stokes-Mueller formalism**
For partially depolarized states, the disordered motions of the electric field **E** of the beam in the xy plane can be properly described only by their statistical properties rather than their instantaneous values. For this reason it is preferable to use field intensities instead of amplitudes. At first sight, one might think that a full probability distribution of the electromagnetic field would be needed to fully characterize such states. In fact, as long as only intensity measurements can be performed with state-of-the-art detectors at optical frequencies, all that is needed to predict the result of any classical measurement are the second moments (quadratic quantities) of the electric field distributions. More specifically, in the framework of linear optics, it can be shown that any possible partially polarized field can be fully characterized by a four dimensional vector, called Stokes vector S, which is defined for any set of orthogonal axes (x, y) as:

$$\begin{pmatrix} I \\ Q \\ U \\ V \end{pmatrix} = \begin{pmatrix} I_x + I_y \\ I_x - I_y \\ I_{45°} - I_{-45°} \\ I_L - I_R \end{pmatrix} = \begin{pmatrix} \langle E_x E_x^* + E_y E_y^* \rangle \\ \langle E_x E_x^* - E_y E_y^* \rangle \\ \langle E_x E_y^* + E_y E_x^* \rangle \\ i\langle E_x E_y^* - E_y E_x^* \rangle \end{pmatrix} \qquad (3)$$

where $I_x$, $I_y$, $I_{+45}$, $I_{45}$ are the intensities which would be measured through ideal linear polarizers oriented along the x, y, +45° and –45° in the plane perpendicular to the propagation direction, while $I_L$ and $I_R$ would be the intensities transmitted by left and right circular polarizers[6]. The Stokes vector is thus defined in terms of directly measurable intensities, which is not the case for the electric field amplitudes involved in the Jones formalism. For fully polarized states, the Stokes vector components are simply given by:

$$\begin{pmatrix} I \\ Q \\ U \\ V \end{pmatrix} = \begin{pmatrix} I_x + I_y \\ I_x - I_y \\ I_{45°} - I_{-45°} \\ I_L - I_R \end{pmatrix} = \begin{pmatrix} E_x E_x^* + E_y E_y^* \\ E_x E_x^* - E_y E_y^* \\ E_x E_y^* + E_y E_x^* \\ i(E_x E_y^* - E_y E_x^*) \end{pmatrix} \quad (4)$$

without any need to average. Conversely, in the most general case of partially polarized light, the brackets on the right hand side of Eq. 3 stand for all possible ways to take averages; e.g., spatially, spectrally or temporally, depending on the sample and measurement conditions. Thus, partially polarized states can be viewed as *incoherent superposition of fully polarized states with different polarizations.*

Within the Stokes formalism, the *degree of polarization* $\rho_S$ related to a given Stokes vector **S** is defined as:

$$\rho_S = \frac{\sqrt{Q^2 + U^2 + V^2}}{I} \quad (0 < \rho_S \leq 1) \quad (5)$$

whereby this quantity varies between 0, for totally depolarized (fully disordered) states, and 1, for totally polarized states.

As the Stokes vector is directly related to intensities, it would undergo a linear transformation upon interaction with a sample. This is described through a 4x4 real matrix **M** called the Mueller matrix[6-8]:

$$\mathbf{S}^{out} = \mathbf{M} \mathbf{S}^{in} \quad (6)$$

Due to the capability of Stokes vectors to describe *any* polarization state, the Mueller matrix can fully describe the polarimetric properties of *any* sample, be it depolarizing or not. In other words, Mueller polarimetry is the only technique able to characterize any sample, under any measurement conditions.

In contrast with the Jones matrix, the Mueller matrix does not carry any information about the overall optical phase shift introduced by the sample. So, depending on whether the overall transmission (or reflectivity) of the sample is of interest or not, the Mueller matrix may be considered in its original or normalized form. In the latter case, its upper left element $M_{11}$ is set equal to 1.

It is important to note that, *while in principle any 2x2 complex matrix may be an acceptable Jones matrix, a real 4x4 matrix is not necessarily a physically realizable Mueller matrix*: a clearly necessary condition is that any acceptable Stokes vector (i.e. with its degree of polarization between 0 and 1) must be able to be transformed into another acceptable Stokes vector. However, this condition alone is not sufficient, and another criterion can be defined from the so-called *coherency matrix* **N**, which is related to the Mueller matrix **M** of interest. Specifically, **M** is physically acceptable if and only if **N** is positive semi-definite; i.e., its eigenvalues are non-negative[9-10]. A sample can be considered as *non-depolarizing* (a condition which depends on the sample but also on the polarimeter used to characterize it!) if and only if its Mueller matrix can be derived from the sample Jones matrix, and the associated coherency matrix **N** exhibits only one strictly positive eigenvalue (whereas the others vanish).

Another criterion, much easier to implement, has also been proposed[11] based on the quadratic depolarization index *P*:

$$P = \sqrt{\frac{\sum_{i,j} M_{i,j}^2 - M_{1,1}^2}{3 M_{1,1}^2}} \qquad (7)$$

which varies from 0 for a perfect depolarizer (only $M_{11}$ is non-zero) to 1 for non-depolarizing matrices.

**Standard Ellipsometry**
Ellipsometry is a well-established and powerful optical tool for the measurement of thin films. Spectroscopic ellipsometry is often used for determining the dielectric functions of various substrates or multi-layered materials. Standard ellipsometric measurements are commonly performed in external configuration, which means that a light beam propagating in air (or vacuum) is reflected by, or transmitted through a sample, and then it propagates again in air (or vacuum) before arriving at the detector. The interest of ellipsometry is that it can measure simultaneously the modulus and phase of the polarization components of the light. The sensitivity of phase measurements, exploited to determine thin film thickness, has its roots in an interferometric effect. The light reflected by the first interface of a layer present in the sample, interferes with the light reflected by the second interface of the layer. The same principle remains valid when a stack of multi-layers are present. Therefore, the maximum film thickness that can be measured with ellipsometry has to be less than the coherence length of the light source. The most common light sources used in ellipsometry are thermal (halogen) bulbs or high pressure arc-discharge plasma lamps, which produce a beam of non-polarized light with spectral frequencies distributed continuously over a broad range, from the ultraviolet (typically 250 nm), to the near infrared (around 2500nm). Accordingly, ellipsometry is capable of characterizing transparent or low absorbing thin films with thickness ranging from less than a nanometer to several micrometers. The use of coherent sources, such lasers, may increase considerably the maximum film thickness measurable (up to several centimeters) at the expense of the spectral bandwidth accessible. There is, of course, the possibility of using lasers working in the super-continuum configuration or particle accelerators (synchrotrons), which provide beams having both large coherence lengths and relatively broad spectral ranges. In spite of those clear advantages, the available facilities in the world for such beams are scarce and accessible only to a restricted number of users and/or applications.

In addition to high sensitivity, ellipsometry has the advantages of being non-destructive and contactless. A spectroscopic ellipsometer is relatively easy to use and requires no sample preparations. Standard ellipsometers can be built with light-weight optomechanical components, and they are relatively compact. They can be mounted as stand-alone instruments or coupled to other systems such as vacuum chambers, chemical reactors or bio-reactors, etc. In the former case, measurements are said to be ex-situ*,* and in the latter they are called in-situ*.* In-situ measurements are interesting because they allow for the characterization of a sample in "real-time" and under the same conditions as it is prepared, deposited or treated (i.e., with no alterations by the atmosphere.)

Historically, ellipsometry has been used to characterize bulk materials, liquids, the surfaces of solids, and multi-layered thin films. The variety of samples that can be studied opens a wide range of possibilities for ellipsometry. A recent survey[12] based on the most relevant database of scientific articles and publications suggests that ellipsometry has been successfully applied in many studies concerning material science (e.g., semiconductors and photovoltaics), biology (biofilms and biosensors) and pharmaceuticals, etc.

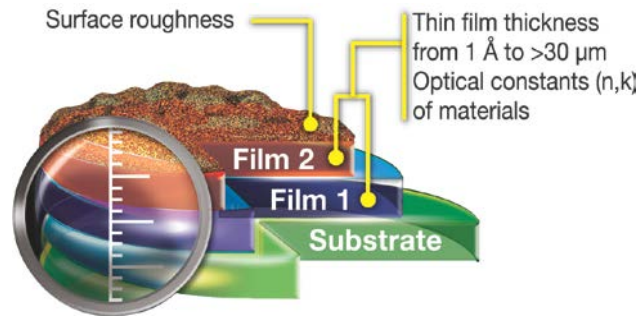

Figure 3. Schematic representation summarizing the different parameters related to the sample that can be deduced using ellipsometry. These parameters include: thin film thickness, optical constants, roughness, porosity, composition, uniformity, etc.

The information provided by ellipsometry is very rich when it comes to layer stack descriptions. It enables accurate measurements of surface roughness and interfaces, while the determination of complex refractive index gives access to fundamental physical parameters which are related to a variety of sample properties including: morphology, crystallinity, chemical composition and electrical conductivity, etc.

Information extracted from an ellipsometric measurement is greatly enhanced by using wavelengths over a wide spectral range, from vacuum ultraviolet to mid-infrared. The far ultraviolet is the most sensitive to small changes such as ultra-thin layers or interfaces, films with low index contrast, gradient and anisotropy. Ultraviolet is also highly sensitive to surface roughness. The near-infrared (NIR) spectral range is necessary to determine the thickness of materials which are strong in absorbing in the visible spectrum. NIR is also used to determine the optical conductivity (typically metals or doped oxides) because in this spectral region the optical response of samples is dominated by free charge carriers.

Because spectroscopic ellipsometry measures two physical magnitudes at each wavelength, the technique obtains more information than standard optical reflection techniques. This capability makes spectroscopic ellipsometry the most accurate thin film measurement tool available.

**Instrumental Implementations of Ellipsometers**
Very many optical configurations can be envisaged for standard ellipsometers. As an exhaustive review of all these designs is clearly beyond the scope of this article, in the following we will restrict ourselves to the configurations schematized in fig. 4. The ellipsometers concerned by the scheme in the figure are made of two optical arms and a sample-holder in between. The first arm, at the entry, comprises a Polarization State Generator (PSG) coupled to a source of light. In all cases the PSG includes a linear polarizer set at an azimuth $P$ with respect to the $p$ direction respect to the plane of incidence. The second arm, or exit arm, is used to determine the polarization of the outcoming beam. It comprises a Polarization State Analyser, or PSA, and a detector which may be a single channel device (photodiode, photomultiplier…) or a multichannel one (typically a CCD coupled with a spectrometer, or, less frequently, with an imaging system). The PSA typically includes a polarizer and possibly other components. The PSG and PSA design actually define the various types of instruments outlined in this part.

Of course, in all cases the polarization components can be inverted: all the PSAs described in the following can be placed in the input.

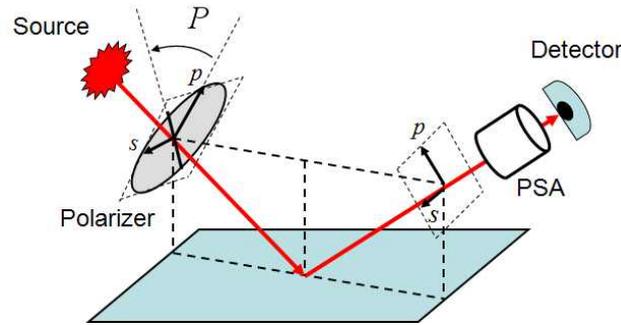

Figure 4. General scheme of a standard ellipsometer. The PSA is the Polarization State Analyzer, which distinguishes the various optical configurations described in this section.

Basically, standard ellipsometers can be classified into two general families, null-ellipsometers and non-null ellipsometers. In null ellipsometers, the optical components of the system must be rotated until the detected intensity vanishes, then the ellipsometric values are deduced from the orientations of the optical elements needed to achieve the null intensity. Conversely, in non-null ellipsometers the light intensity is modulated temporally by the action of at least one of the optical components integrating the ellipsometer, then after an harmonic analysis of the signal, the ellipsometric values are deduced. The non-null ellipsometers can be classified into three groups: rotating polarizers or analyzers, rotating compensators and phase-modulated. In the following we overview some characteristics of the different types of ellipsometers.

*Null-ellipsometers*
Null ellipsometers were the first type of instruments to be developed in late nineteenth century because of their instrumental simplicity and ease of use. In the former systems, rotation of the optical elements to achieve the null intensity was done manually and the null was evaluated with the naked eye. During the twentieth century, thanks to the generalization of electronics, automatic rotation by motors and photodiodes substituted the human hand and eye respectively, making the measurement task much more comfortable. The basic PSA of a null ellipsometer is made of a quarter wave plate and a rotatable polarizer. The fast axis of the quarter wave plate is placed at 45° respect to the direction parallel to the plane of incidence. The intensity measured by the detector is then:

$$I = \sin(2A)\sin(2\Psi)[\sin(2P)\cos(\Delta) - \cos(2P)\sin(\Delta)] - \cos(2A)\cos(2\Psi) + 1 \quad (8)$$

Which vanishes only if $A = \Psi$ and $2P + 90° = \Delta$; (9) meaning that now the ellipsometric angles $\Psi$ and $\Delta$ are retrieved from the orientations of the input polarizer $P$ and the output analyzer $A$. Null ellipsometers based on quarter-wave retarders have been shown to be very accurate, and comparable to good modern instruments, but limited to a single wavelength due to the dispersion of the retarder. To circumvent this limitation a variable retarder such a Babinet Soleil or another equivalent optical device, can be used to make the instrument spectroscopic. However, the overall accuracy may be limited by that of the variable retarder calibration.

*Rotating polarizer – analyzer ellipsometers*
Rotating analyzer or polarizer ellipsometers use two polarizers (polarizer and analyzer), and one of them is continuously rotating. This simple mechanical rotation is used to harmonically modulate the intensity of the light for subsequent synchronous detection. When the analyser is rotated, the optical configuration is often referred as "PSRA" for Polarizer-Sample Rotating Analyzer. Conversely, when the polarizer is rotated the configuration is called "RPSA". The detected signal by a PSRA ellipsometer can be written as follows:

$$S(t) = S_0 [1 + \alpha \cos(2\omega t) + \beta \sin(2\omega t)] \qquad (10)$$

Where $\omega$ is the angular rotation speed of the analyzer. The Fourier coefficients of the modulated signal can be written as functions of the ellipsometric angles $\Psi$, $\Delta$ and the orientation of the polarizer with respect to the plane of incidence, $P$:

$$\alpha = \frac{\tan^2 \Psi - \tan^2 P}{\tan^2 \Psi + \tan^2 P}, \quad \beta = \frac{2 \tan \Psi \cos \Delta \tan P}{\tan^2 \Psi + \tan^2 P} \qquad (11)$$

from which one easily gets

$$\tan \Psi = \sqrt{\frac{1+\alpha}{1-\alpha}} |\tan P| \quad \cos \Delta = \frac{\beta}{\sqrt{1-\alpha^2}} \times \frac{\tan P}{|\tan P|} \qquad (12)$$

As a result tan$\Psi$, and thus $\Psi$ itself, is determined unambiguously. In contrast, *as only* cos$\Delta$ *is actually retrieved*, for this type of instrument :
- Only the absolute value of $\Delta$ is measured,
- This value becomes inaccurate when $\Delta$ is close to 0 or 180°, where the cosine function reaches its extrema. This situation typically occurs for thick transparent or highly absorbing samples.

However, this shortcoming may be obviated by inserting an additional known retarder, with its axes aligned with the *s* and *p* directions, to "shift" the retardation to be measured away from 0° or 180°. Another possible issue to be solved are the systematic errors which may be introduced by any residual polarization of the source and/or of the detector. On the other hand, as the technique uses only polarizers, it is possible to operate it over wide spectral ranges (from 200 nm to 30 µm), and the rotation speed may be chosen according to other requirements, such as a possible acquisition by a linear CCD after a spectrometer, which can be very convenient in many cases. Concerning the Muller matrix elements, it can be shown[13] that even in the most favorable configuration, the element $M_{44}$, is not accessible. When the PSA at the output consists in a simple (rotating) linear analyzer without a compensator, the fourth row of the Mueller matrix is also inaccessible. Only the upper left 3x3 sub-matrix of the sample Mueller matrix can be determined, provided the measurements and data analysis outlined above are repeated with at least four different azimuths *P* of the input polarizer.

*Rotating compensator ellipsometers*

Rotating compensator ellipsometers include at least one linear retarder, usually called (somewhat improperly) compensator. Depending on whether the rotating compensator is placed at the entry or at the exit arm there are two possible configurations known as PRCSA or PSRCA where the meaning of *P*, *S* and *A* is the same as previously RC stands for Rotating Compensator. In the following we will consider the PSRCA. A major difference between rotating compensator and rotating analyzer ellipsometers, is that with a rotating compensator and a fixed linear analyzer it is possible to retrieve all four components of the Stokes vector **S**$_{out.}$, implying that more quantities are measurable, both in standard ellipsometry and for Mueller matrices. If the compensator is a quarter wave plate (retardation equal to 90°), the intensity recorded by the detector in the PSRCA configuration can be written as follows:

$$I = I_0 \left(2 - \cos 2\Psi + 2\sin 2\Psi \sin \Delta \sin 2C - \cos 2\Psi \cos 4C + 2\sin 2\Psi \cos \Delta \sin 4C \right) \quad (13)$$

where $I_0$ is the non-modulated (DC) intensity provided by the source, and $C = \omega t$, the compensator orientation which of course, depends on time. As a result, the three different Fourier harmonics of the modulated signal directly provide the three quantities $\cos 2\Psi$, $\sin 2\Psi \sin \Delta$ and $\sin 2\Psi \cos \Delta$. In other words, rotating compensator ellipsometers provide accurate measurements of the ellipsometric $\Psi$ and $\Delta$ angles over the complete measurement range ($\Psi$=0-90°; $\Delta$=0-360°). Similar results can be obtained for PRCSA ellipsometers.

However, the construction of a rotating compensator ellipsometer, with a compensator which behaves ideally providing an achromatic retardance of 90° over a wide spectral range, is a difficult optomechanical challenge, and it requires more complicated calibration and data reduction procedures than rotating polarizer or analyzer ellipsometers. Any deviation of the optical response of the compensator from the ideal behavior must be carefully calibrated, otherwise it will be the source of important systematic errors. Rotating compensator ellipsometers can be implemented in more general configurations, among which :
   a) *The RP/RCFA configuration*, which consists of a rotating polarizer at the entry arm and a rotating compensator followed by a fixed analyzer at the exit arm.
   b) *The FPRC/RA configuration*, which consists of a fixed polarizer and rotating compensator at the entry arm and at rotating analyzer at the exit arm.

Those configurations are often used when it comes to determine partial Mueller matrices instead of ellipsometric angles. In the best operation mode of the RP/RCFA configuration, the compensator and the polarizer are rotated synchronously at different frequencies. In an optimal operation configuration the rotation frequency of the polarizer is 3 times the one of the compensator. Then the detected signal can be decomposed in a Fourier series:

$$4\frac{I}{I_0} = \alpha_0 + \sum_{j=1}^{7} \left(\alpha_{2j} \cos(2jP) + \beta_{2j} \sin(2jP)\right) \quad (14)$$

where $I_0$ is the light source intensity. The Fourier analysis of the modulated signal provides 15 coefficients which allow to determine the elements of the first three columns of the Mueller matrix as follows:

$$\mathbf{M} = \begin{pmatrix} (\alpha_0 - \alpha_6) & (\alpha_1 - \alpha_5 - \alpha_7) & (\beta_1 - \beta_5 + \beta_7) & \bullet \\ 2\alpha_6 & 2(\alpha_5 + \alpha_7) & 2(\beta_7 - \beta_5) & \bullet \\ 2\beta_6 & 2(\beta_7 + \beta_5) & 2(\alpha_5 - \alpha_7) & \bullet \\ -2\beta_3 & -2\beta_2 & -2\alpha_2 & \bullet \end{pmatrix} \quad (15)$$

The Fourier components are functions of the compensator properties, in particular retardation, which can be wavelength dependent. The calibration of such as system is extremely complex, especially when the ellipsometer is spectroscopic[14]. Conversely, the advantage of such as system is that a single measurement scheme allows to obtain 12 out of 16 Mueller matrix elements. If a simplified operation mode is used, in which only the compensator is rotated continuously, the 12 elements of the Mueller cannot be obtained after a single measurement. The polarizer must be placed at different azimuths, and for each position, a new measurement must be made. Once the process is finished, the combination of the Fourier coefficients extracted from all the measurements allows to obtain the first three columns of the Mueller matrix. Similar arguments can be given to illustrate the operation of the FPRC/RA configuration which then provide the first three *rows* of the Mueller matrix.

*Phase-modulated ellipsometers*
Finally, there are the phase-modulated ellipsometers which include at least one photo-elastic modulator (PEM). The first phase-modulated ellipsometer was built in the mid XXth century[15], since then it has reached a considerably popularity. Two types of phase-modulated ellipsometer are commercialised by HORIBA Scientific under the name of UVISEL and UVISEL2. In a phase-modulated ellipsometer, the PEM can be placed between the linear polarizer and the sample, either at the entry or exit optical arm, giving rise to the PMSA or the PSMA configurations respectively. Here *P*, *M*, *S* and *A* stand for fixed polarizer, modulator, sample, and fixed analyzer respectively. How does the PEM work? The PEM consists of a bar of a transparent material (typically fused silica) exhibiting isotropic behavior when no mechanical stress is applied on it. Once mechanical stress is applied, the bar becomes birefringent, which means that light travels faster along one optical axis than along the other when passing through the bar. Birefringence produces a different phase velocity for each component of the polarized beam, and a modulated phase shift is therefore induced. The optical response of the bar can be described as a non-uniform biaxial medium.

Mechanical stress is usually applied using piezoelectric transducers attached to the end of the bar. The transducers are not static but rather vibrate at a given frequency, which in turn produce an oscillating stress. The frequency is selected to be close to a mechanical resonance of the bar in order to enhance the effect of the transducers. For silica bars of several centimetres in length, the resonance frequency is close to 50 kHz. The high modulation frequency provides signal measurements in a wide dynamic range with low noise level. When combined with powerful digital signal averaging and highly sensitive detectors, phase-modulated ellipsometers provide the best signal-to-noise ratio from vacuum ultraviolet (VUV) to NIR, as well as the most sensitive measurements. The interested reader can find information about other types in different monographs[1-4].

We now consider the PSMA configuration in which at the entry arm the polarizer is fixed and set at an azimuth *P* with respect to the plane of incidence, while in the exit arm, as represented in fig. 4, the photoelastic modulator is set to an azimuth *M* and the linear analyzer is set at an azimuth angle *A* with respect to the plane of incidence. The detected signal then takes the form :

$$S(t) = S_0 [1 + I_s \sin(\delta(t)) + I_c \cos(\delta(t))], \quad (16)$$

with:

$$I_c = \sin[2(A-M)] [\sin 2M (\cos 2\Psi - \cos 2P) + \sin 2P \cos 2M \sin 2\Psi \cos\Delta] \quad (17)$$

$$I_s = \sin[2(A-M)] \sin 2P \sin 2\Psi \sin\Delta$$

In practice, as $\delta(t) = \sin(\omega t)$, the preceding expressions must be developed in Fourier series (with the well-known Bessel functions as coefficients) to express $I_s$ and $I_c$ as functions of the directly measured quantities, actually the amplitudes of the $\sin(\omega t)$ and $\sin(2\omega t)$ components of the signal. It can be shown that the signal S (*Is, Ic*) is maximized when $A-M = 45°$. Moreover it is also clear that it is not possible to unambiguously determine $\Psi$ and $\Delta$ from a single measurement configuration. In practice, two configurations are typically used

- $M=0°$, $A=45°$, $P=45°$, known as configuration II, for which we get $I_s = \sin 2\Psi \sin\Delta$, $I_c = \sin 2\Psi \cos\Delta$

- $M=45°$, $A=90°$, $P=45°$, known as configuration III, for which we get $I_s = \sin 2\Psi \sin\Delta$, $I_c = \cos 2\Psi$

In practice, all that is needed to shift from one configuration to the other is to rotate the PSA, which can be done automatically without major difficulties, and then combine the results of the two measurements for a complete, unambiguous determination of both $\Psi$ and $\Delta$. For Mueller matrix measurements, the three quantities which can be directly retrieved from the Fourier Analysis of the signal can be recast in terms of the matrix elements $M_{ij}$ and the azimuths $P$, $A$ and $M$ as[13,16].

$$\begin{aligned}
I_\alpha &= S_0 = M_{11} + M_{12}\cos(2A) + M_{13}\sin(2A) \\
I_\beta &= S_0 I_S = (M_{31} + M_{32}\cos(2A) + M_{33}\sin(2A))\cos(2M) - \\
&\quad - (M_{21} + M_{22}\cos(2A) + M_{23}\sin(2A))\sin(2M) \\
I_\gamma &= S_0 I_C = M_{41} + M_{42}\cos(2A) + M_{43}\sin(2A)
\end{aligned} \qquad (18)$$

from which it is clear that the nine elements $M_{1i}$, $M_{2i}$ and $M_{4i}$ can be retrieved with three measurements carried out with $M= 45°$ and $A = 0°$, $60°$ and $120°$ for example. Then the last four element $M_{3i}$ can be obtained by another three measurements, with the same $A$ values as before but $M = 0°$. Moreover, this new set of measurements over-determines the values of $M_{1i}$ and $M_{4i}$. As a result, six measurements are necessary to retrieve the full set of 12 elements of the first three columns of the Mueller matrix, with partial redundancy.

In this subsection we have presented the most commonly used experimental configurations for standard ellipsometry, with particular emphasis of the quantities that actually can, or cannot, be measured by each of them. In table I, we summarize the main characteristics of these configurations, including their strengths and weaknesses.
We want to make two points absolutely clear :
- We did not try to review the many refinements are more complex systems which have been tested and developed, possibly up to commercialization.
- By listing the main advantages and weaknesses of each technique we absolutely do not mean that commercially available systems using this technique necessarily presents these strengths and weaknesses. While some basic limitations, such as those concerning the measurable Mueller matrix elements, cannot be solved in a given configuration, many other practically essential issues, among which those related to the measurements accuracy, the speed, the signal to noise ratio and the like greatly depends on engineering developments which are clearly beyond the scope of this contribution.

We thus stress that the information presented in Table I is by no means a "buyer's guide": it might be useful only to ask the manufacturers some reasonably relevant questions !

**Analysis of ellipsometric data**

Conventional techniques used for thin film characterizations (e.g., ellipsometry and reflectometry) rely on the fact that the complex reflectivity of an unknown optical interface depends on both its intrinsic characteristics (material properties and thickness of individual layers) and on three properties of the light beam that is used for the measurements: wavelength, angle of incidence, and polarization state. In practice, characterization instruments record reflectivity spectra resulting from the combined influence of these parameters. The extraction of the information concerning the physical parameters of the sample from the recorded spectra is an indirect process. In other words, from a given ensemble of experimental data, we are interested in building a theoretical model of the sample which we are hoping to reproduce as closely as possible to the measured data. In general, theoretical models depend on a series of parameters characteristic of the sample, which must be adjusted to make the theoretical data "fit" the measurements. A common model for a stack of layers includes the thickness and refractive indexes of all the layers. In many cases, the refractive index of the substrate must be considered, as well[17-18]. The quality of the fit is usually evaluated with a figure of merit, and it is used during the fitting process to guide the numerical algorithm which searches for the best-fitted values of the model parameters. According to Jellinson[17], it is necessary to define an unbiased figure of merit in order to judge how well the data fit. There exist different expressions for the figure of merit, but the most popular is the one based on the mean square root of the differences between simulated and measured data.

$$\chi^2 = \frac{1}{N - M - 1} \sum \frac{\left(\Psi^T - \Psi^{Exp}\right)^2}{\sigma_\Psi^2} + \frac{\left(\Delta^T - \Delta^{Exp}\right)^2}{\sigma_\Delta^2} + \frac{\left(R^T - R^{Exp}\right)^2}{\sigma_{R'}^2} \qquad (19)$$

N refers to the total number of data points and M is the total number of fitted parameters. The superscripts "T" and "Exp" refer to the theoretical and experimental data, respectively. The summation is done over all the spectral data points. The sigmas in the denominators correspond to the estimated uncertainty in the experimental values. Typical values for sigmas of HORIBA Scientific ellipsometers are around 0.5 and 0.1%. The advantage of the formulation above is that it allows us to include non-ellipsometric data, such as total reflectivity R, in the fitting process. The combination of ellipsometric data with information coming from other sources can be interesting, and enhances accuracy in the determination of fitted parameters. According to Jellison[17] the figure of merit behaves like a multivariate mathematical function which depends on a given number of fitting parameters.

Once the figure of merit has been defined, it is possible to take advantage of computers to automate the fitting process, which is based on the search of the minimum value of the figure of merit. The automatic process of minimization of a multivariate function is far from trivial. The principal difficulty that arises almost systematically is the fact that the figure of merit may have either multiple minima with the same value, or multiple partial minima with different values. In order to minimize the impact of this drawback in the final results, it is preferable to use smart or advanced minimization strategies which are based either on systematic multiple guesses for the initial parameters, genetic algorithms, or even simulated annealing algorithms. Despite the advantages of those minimization strategies, it is important to note that at the end of a minimization process, review of the results is necessary to check pertinence, accuracy and efficiency.

A second factor that can complicate data fitting, which is inherent to the fact that ellipsometry data analysis is an indirect process, is the correlation between fitted parameters. We talk of parameter correlation when it is possible to find multiple sets of parameters that produce the same value of the figure of merit. Correlation is said to be linear when the couples of correlated parameters follow a linear relation. Correlation between fitting parameters happens because experimental data are not sensitive to individual parameters, but to a combination of them. Correlation between two parameters may also occur if one of the two parameters has much more impact on the data (i.e., the optical response) than the other. Correlation is specific to the studied sample and therefore it is difficult to establish general rules to treat the problem. However, whenever correlation appears, it is advised to keep one of the parameters fixed to a 'reasonable' value, which can be obtained from a complementary technique (e.g., TEM, XPS…), and fit the rest.

HORIBA Scientific's instruments, either ellipsometers or polarimeters, come with modelling software, DeltaPsi2, which has been specially designed to help the user to overcome the most commonly encountered difficulties in data analysis. DeltaPsi2 provides different fitting procedures based on multi-start or multi-guess strategies to avoid problems related to multiple partial minima. DeltaPsi2 also provides a statistical analysis of the fitting procedure to detect and evaluate possible correlations among the fitted parameters. The statistical analysis of data is based on the variance-covariance matrix formalism[17]. DeltaPsi2 software has an easy to use graphical user interface and it has become a critical reference tool among ellipsometrists and non-specialized users as well.

Despite the great advantages of ellipsometry, it is limited to the analysis of samples which do not depolarize light. As stated previously, depolarization arises because of the incoherent superposition of light with different polarization states. In practice, depolarization is commonly encountered when measuring inhomogeneous samples (either in terms of composition or thickness), or very rough surfaces. When depolarization is present, it is no longer possible to use ellipsometry and the related optical models. In such cases it is necessary to measure with a Mueller polarimeter and use advanced optical models to take into account the depolarization. A more complete discussion of Mueller Ellipsometry will be given in the next chapter.

**Examples of Accuracy of Ellipsometric Measurements**
The sensitivity of ellipsometry can be quantified by the simulated effect of the presence of an ultra-thin layer on the two measurable ($\Psi$, $\Delta$). The table included in fig. 5 below shows the calculated results corresponding to a substrate of crystalline silicon (c-Si) with n=3.8819 and k=0.019 at 633nm, coated with a transparent film of silicon dioxide ($SiO_2$) with n=1.5 and k=0. Under these conditions, it is seen that $\Delta$ changes by about 0.3° and $\Psi$ by 0.001° per 1 Å of film. These results also show that $\Delta$ is the most sensitive parameter to small changes as it varies by 2.976° for 10 Å, versus 0.015° for $\Psi$.

Considering that a properly aligned ellipsometer with high quality optics is capable of precision of about 0.01-0.02° in $\Delta$ and $\Psi$, then a theoretical sensitivity to thickness on the order of 0.01nm is achievable thanks to $\Delta$. And since an atomic layer thickness is on the order of 0.1nm, it is then possible to conclude that the high sensitivity of ellipsometry allows users to detect changes of one monoatomic layer in the thickness of a silicon oxide layer.

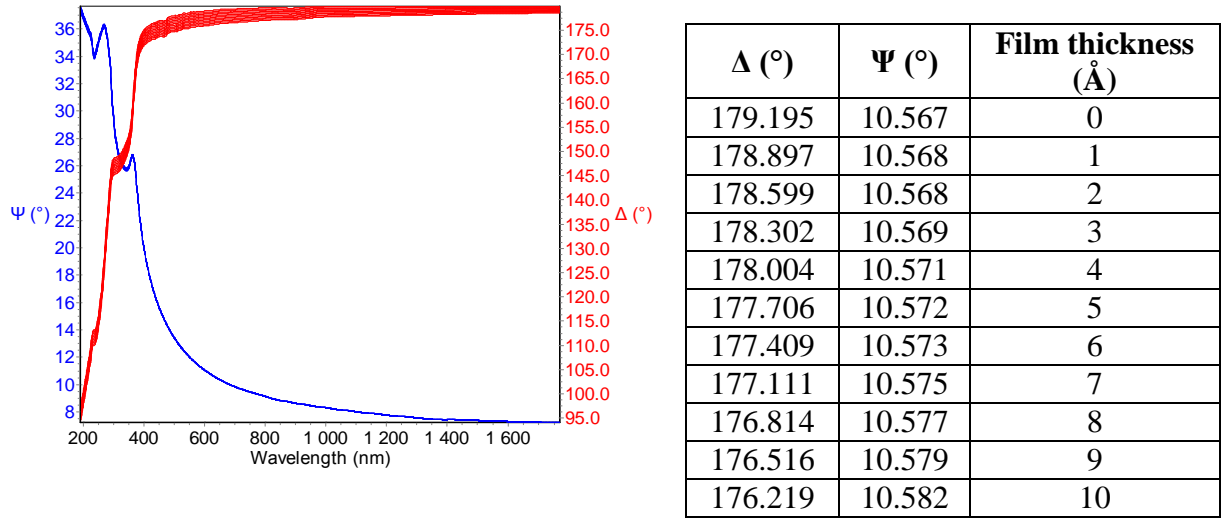

| Δ (°) | Ψ (°) | Film thickness (Å) |
|---|---|---|
| 179.195 | 10.567 | 0 |
| 178.897 | 10.568 | 1 |
| 178.599 | 10.568 | 2 |
| 178.302 | 10.569 | 3 |
| 178.004 | 10.571 | 4 |
| 177.706 | 10.572 | 5 |
| 177.409 | 10.573 | 6 |
| 177.111 | 10.575 | 7 |
| 176.814 | 10.577 | 8 |
| 176.516 | 10.579 | 9 |
| 176.219 | 10.582 | 10 |

Figure 5. (Left). Simulated Ψ (blue) and Δ (red) angles over the spectral range from 200 to 1700nm, for a c-Si substrate covered with a thin layer of $SiO_2$. The different spectra (clearly seen in Δ) correspond to different thickness of the $SiO_2$ layer.
(Right). Table II. The different thickness values together with the calculated Ψ and Δ angles at 633nm wavelength.

In this way, it is also important to point out that the power of spectroscopic measurement improves the precision of thickness determination. Spectroscopic measurements means being able to measure (Ψ, Δ) at each wavelength. In the case of a layer-covered substrate, the general formula for ellipsometry relates measurements to properties as follows: (Ψ, Δ) = f($\varepsilon_0$, $\varepsilon_1$, $\varepsilon_2$, d, λ, θ), where λ is the wavelength of light, θ the angle of incidence, d the film thickness, $\varepsilon_0$ optical properties of air, $\varepsilon_1$ optical properties of the layer, and $\varepsilon_2$ optical properties of the substrate. λ, θ, $\varepsilon_0$ and $\varepsilon_2$ are known, therefore from a measurement of (Ψ, Δ), two properties can be obtained generally: d and n (refractive index). Hence, spectroscopic ellipsometric measurements enable the determination of thickness (d) at each (Ψ, Δ) couple improving its precision.

When the layer becomes very thin, or in the case of very thin interfaces or films with low index contrast, the technique provides the best sensitivity in the VUV wavelength range. The example below illustrated in fig. 6 shows the variations of (Ψ, Δ) over the spectral range of 200-1700nm for a glass substrate, Corning 1737, covered with a $SiO_2$ layer varying from 0 to 10 nm by a step of 1nm. The tab. III compares the (Ψ, Δ) values at 633nm and 190nm for 0 and 10 nm $SiO_2$ layer.

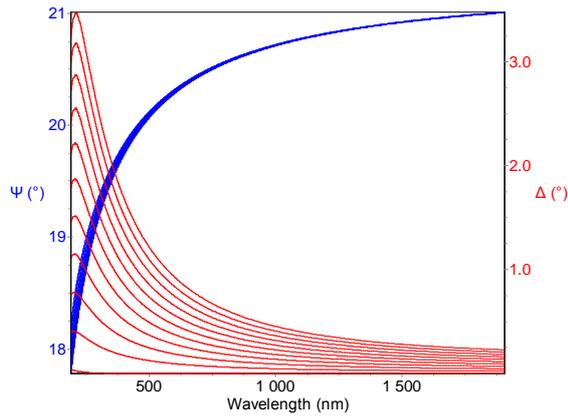

| Thickness (nm) | λ 190 nm | | λ 633 nm | |
|---|---|---|---|---|
| | Ψ | Δ | Ψ | Δ |
| 0 | 17.78 | 0.038 | 20.34 | 0.001 |
| 10 | 18.20 | 3.233 | 20.37 | 0.861 |
| Diff. | 0.422 | 3.195 | 0.03 | 0.86 |
| Absolute error | 0.01 | 0.01 | 0.005 | 0.005 |
| Diff / Error error | 42 | 320 | 6 | 170 |

Figure 6. Simulated Ψ (blue) and Δ (red) values over the spectral range from 200 to 1700 nm, for a glass substrate covered with a thin film of SiO2. The different spectra correspond to different thicknesses of the SiO$_2$ layer varying from 0 to 10 nm by a step of 1 nm.

Table III. Calculated Ψ and Δ vales at 633nm and 190 nm for 0 and 10 nm SiO$_2$ layer. The table also includes an estimation of the absolute experimental uncertainty of Ψ and Δ at 633nm and 190nm. The ratio between the differences among the Ψ and Δ measured at a thickness of 0 and 100 nm respect to the uncertainty in the measurements, give an idea of the sensitivity of Ψ and Δ at different wavelengths.

From these results, it is obvious that the phase information (Δ) is very sensitive to single layer thickness with a stronger effect in the FUV. This example also shows the importance of accurate Δ measurements around 0°, which are provided only by certain types of ellipsometers, including phase-modulated, rotating compensators, and rotating polarizer/analyzer with additional retarders.

**Mueller Ellipsometry**

Mueller Ellipsometry or Polarimetry, is aimed at characterizing the polarimetric properties of the sample under study, by measuring the polarization changes induced by this sample on selected input polarization states, defined by a Polarization State Generator (PSG). The output polarizations are analyzed by means of a Polarization State Analyzer (PSA) followed by a detector, according to the general scheme outlined in fig. 7. Of course, the incident polarizations defined by the PSG and/or those detected by the PSA may vary during a given experiment.

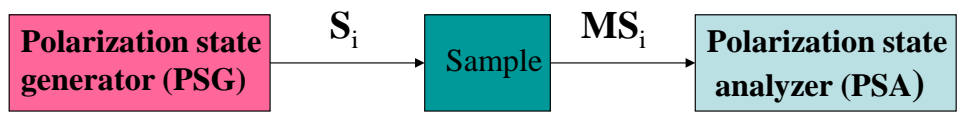

Figure 7. General principle of operation of any Mueller Ellipsometry.

The PSG produces a set of input Stokes vectors $S_i$, which are transformed by the sample into $M\,S_i$ ($M$ being the Mueller matrix of the sample). These output Stokes vectors are then analyzed by the PSA, which delivers the raw signals $B_{ij}$ by projecting each vector $M\,S_i$ onto its basis states. This scheme can be summarized by the simple matrix equation.

$$\mathbf{B} = \mathbf{A}\,\mathbf{M}\,\mathbf{W} \qquad (20)$$

where the modulation matrix **W**, which characterizes the PSG, is formed by the $S_i$ vectors in columns, while the $S'_j$ are the line vectors of the analysis matrix **A** characterizing the PSA. In the most general case, **B** is rectangular, with *m* lines and *n* columns, where *m* and *n* respectively represent the numbers of states generated by the PSG and analyzed by the PSA.

To get the full Mueller matrix **M**, both the PSG and the PSA must be "complete", with at least 4 basis states. Then Eq. 20 is sufficient to extract **M** from **B** by merely inverting the (in principle well known !) matrices **A** and **W**, if both *m* and *n* are equal to 4, or by pseudo-inverting these matrices if the system is overdetermined. In the following for simplicity reasons, we will consider only the case of "minimal" Mueller ellipsometers, for which *m* = *n* = 4, but we emphasize that all the ideas exposed in the following about instrument optimization and calibration can be easily transposed to overdetermined configurations.

While their principle of operation may seem straightforward, Mueller ellipsometers are not so widespread (only two have been very recently made commercially available), because of the added technical complexity due to the simultaneous presence of *complete* PSG *and* PSA. Two issues are of paramount importance (as for any other instruments, but they are particularly critical here)

- The ***optimization of the instrument design***, to get the optimal performance if all components were ideal (perfectly well described by the model). The general criterion for this optimization, namely the minimization of the condition numbers of matrices **A** and **W** is now widely accepted.
- The instrument ***calibration***, in other words the determination of the actual **A** and **W** matrices, which are necessarily affected by the many imperfections of the optical components, positioning systems and the like. Actually, for such complex systems, the usual approach based on a detailed modeling of the whole instrument and its non-idealities may be totally inapplicable. Conversely, the Eigenvalue Calibration Method developed and experimentally validated by E. Compain[19] circumvents this problem by determining both **A** and **W** matrices from a set of measurements on reference sample directly, by algebraic methods, *without any modeling of the instrument.* As a result, no very specific samples, such as retardation plates with accurate retardation values, are needed.

Due to its flexibility and robustness, the ECM has been a cornerstone of all the instrumental developments in Mueller ellipsometry at LPICM (and a few other laboratories as well). Its usefulness could hardly be overestimated for the development of innovative Mueller ellipsometers. Item 1 is probably the easiest to address. If we rewrite Eq. 20 as

$$\mathbf{M} = \mathbf{A}^{-1} \mathbf{B} \mathbf{W}^{-1} \quad (21)$$

we see that the optimization of the instrument design is equivalent to a minimization of the errors in **M** for a given value of the measurement errors in the raw matrix **B**. Due to the algebraic properties of matrices, the error propagation from **B** to **M** will be minimized if the *condition numbers* of **A** and **W** are minimized[20-23]. Without trying to be too rigorous, we now illustrate the rationale behind this criterion by considering the noise propagation from raw intensities to final results in the case of a PSA.

The condition number $c(\mathbf{X})$ of a given square matrix $\mathbf{X}$ is defined as $c(\mathbf{X}) = \|\mathbf{X}\|\|\mathbf{X}^{-1}\|$, (22) where the norm of the matrices (and vectors) can be defined in several ways. In our case, the most relevant choice is the Euclidean norm for the vectors while for matrices we define $\|\mathbf{X}\| = \sup[s_i(\mathbf{X})]$; (23), where $s_i$ are the singular values of $\mathbf{X}$. In the case of a general polarimeter, Eq. 20 can be written as: $\mathbf{B}^{(16)} = [\mathbf{W}^T \otimes \mathbf{A}]\mathbf{M}^{(16)} = \mathbf{Q}\mathbf{M}^{(16)}$; (24), with $\mathbf{M}^{(16)}$ and $\mathbf{B}^{(16)}$, written as 16 component vectors and $\mathbf{Q}$, being a 16 by 16 matrix.

It can be shown that the condition number of Q is the product of the condition number of A and W. $c[\mathbf{W}^T \otimes \mathbf{A}] = c(\mathbf{Q}) = c(\mathbf{W})c(\mathbf{A})$ ;(25). The noise, $\delta M$, on the computed Mueller matrix is directly linked to the noise on the measurement $\delta B$ by the following relationship: $\delta \mathbf{S} = \mathbf{Q}^{-1}\delta \mathbf{B}$ ; (26). By applying the norm on Eqs. 24 and 26, supposing that the measurement noise comes primarily from the matrices $\mathbf{A}$ and $\mathbf{W}$, the relative error on the Mueller matrix is bounded by:

$$\frac{\|\delta \mathbf{M}\|}{\|\mathbf{M}\|} \leq \|\mathbf{Q}^{-1}\| \cdot \|\mathbf{Q}\| \cdot \frac{\|\delta \mathbf{B}\|}{\|\mathbf{B}\|} = c(\mathbf{Q}) \cdot \frac{\|\delta \mathbf{B}\|}{\|\mathbf{B}\|} \quad (27)$$

To minimize the relative errors on $\mathbf{M}$, we have to minimize the condition number of $\mathbf{Q}$. We have to optimize both condition numbers of the PSG and the PSA. The theoretical limit for the condition number of a matrix is 1 when the matrix is unitary. However, the matrices $\mathbf{A}$ and $\mathbf{W}$ are special matrices: their rows (or columns) are Stokes vectors representing totally polarized states which implies some theoretical bounds. The condition number of the matrix of the PSG and the PSA is bounded by √3.

Finally, we point out that the minimization of the condition numbers $c(\mathbf{A})$ and $c(\mathbf{W})$ optimizes the propagation of *additive noise* such as Gaussian noise. In principle, other indicators may be found to minimize the effect of other types of noise, such as the multiplicative noise due to speckle effects in imaging with spatially coherent light. In practice, this criterion provides very efficient guidelines to optimize the design of complete Mueller ellipsometer, as it has been experimentally demonstrated among others, on a double rotating compensator setup operated in discrete rotation steps[24].

In addition to the "Standard" double rotating compensator operated with continuous rotations[25], many optimized designs of complete PSA and PSG have been published in the past decade, based on photoelastic modulator in double pass[26], achromatic division of amplitude prism[27], Pockels cells[28,29] nematic or ferroelectric. The last two types of PSGs and PSAs will be described in more detail below.

Last but not least, we conclude this subsection with the two following remarks
- Minimizing the conditions numbers of $\mathbf{A}$ and $\mathbf{W}$ not only minimizes the noise on the extracted Mueller matrix $\mathbf{M}$, but it also "equalizes" the noise among its various components[22], and is thus recommended only for the complete Mueller ellipsometers described in this section. For more specialized instruments, or when particular attention is paid to some particular elements of $\mathbf{M}$, other criteria may be much more adapted.
- In principle, the minimization of $c(\mathbf{A})$ and $c(\mathbf{W})$ is intended to minimize the effects of statistical noise on $\mathbf{B}$, but in practice it turns out to be also a good criterion to minimize *systematic errors* even though such errors cannot be treated by a general theory comparable to those available for statistical noises.

In the following we briefly outline various widely used configurations for PSGs and PSAs, without trying to be exhaustive. We first consider those based on what we call "traditional" approaches, which make use of the elements previously described for standard ellipsometers, with, however, suitable modifications to provide full Mueller matrix measurements. We then focus on the original systems developed at LPICM, and based on nematic and ferroelectric liquid crystals. These devices are actually extremely easy to use, and typically feature wide angular and spatial acceptances, which make them particularly well suited for imaging applications, in the visible and near infrared range. Spectroscopic Mueller ellipsometers based on these devices have also been successfully developed and commercialized by HORIBA Scientific. For all the optimized PSGs described in the following, the corresponding PSAs are nothing else but the mirror images of the PSGs.

*Traditional approaches*
Two of the standard ellipsometric configurations, the rotating compensator and the photoelastic modulator have been generalized, at the expense of extensive instrumental and calibration complication, in order to access the full Mueller matrix.

Concerning the rotating analyzer configuration, the generalization consists of using at least two rotating compensators, both with an ideal retardance of 90° and rotating synchronously with different angular speeds[14, 24, 25]. One compensator is placed at the entry arm between the polarizer and the sample, whereas the second compensator is placed at the exit arm between the sample and the analyzer. Following the nomenclature previously described, this configuration can be addressed as PRCSRCA, or in a shortened version just as PCSCA. The advantage of this configuration is that it allows to access the full Mueller matrix in a single measurement run. This approach has been used to develop a commercially available spectroscopic Mueller ellipsometer[30].

The second type of generalized ellipsometer, based on photoelastic modulators, can be found in two variants. The first one, similar to the rotating compensator consists of a system with two modulators. One modulator is placed at the entry arm, between the polarizer and the sample. The second modulator occupies a symmetric position respect to the first. It is placed at the exit arm between the sample and the analyzer. This configuration can be called, PMSMA. The two modulators can be operated synchronous or asynchronously, but they must be resonant at different frequencies. The drawback of this configuration is that in order to access the whole Mueller matrix, the modulators must be placed at different orientations, and that a complete measurement must be carried out for each orientation[31]. The second variant, consist of a system with four photoelastic modulators. According to the description given by Arteaga[32], two modulators are placed between at the entry arm between the sample and the polarizer, and two modulators are placed at the exit arm between the sample and the analyzer. Again the modulators must vibrate at different frequencies in order to get maximum sensitivity and to avoid possible ambiguities. The advantage is that the four-modulator configuration is exempt of mechanical movements and therefore it can measure the full Mueller matrix in a single run.

*Nematic Liquid Crystals*
These devices behave as electrically controllable variable retarders, analogous to Babinet Soleil Bravais compensators, with fixed orientation of their slow and fast axes and retardations which may be adjusted from 1~2 times 360° to almost 0° by applying a.c. driving voltages, typically in square wave form, with rms values from 0 to about 15 V. We used nematic liquid crystal (NLC) variable retarders from Meadowlark; detailed information about these devices is available on their site[33]. One limitation of NLCs is their slow switching times, of the order of tens of milliseconds.

The whole PSG is composed of a linear polarizer followed by two NLCs with their fast axes set at the (fixed) azimuths $\theta_1$ and $\theta_2$ with respect to the polarization defined by the polarizer. Calling respectively $\delta_1$ and $\delta_1$ the retardations of the NLCs, a straightforward calculation provides the output Stokes vector

$$\mathbf{S}_{PSG} = \begin{pmatrix} 1 \\ (c_1^2 + s_1^2 \cos\delta_1)(c_2^2 + s_2^2 \cos\delta_2) + c_1 c_2 s_1 s_2 (1 - \cos\delta_1)(1 - \cos\delta_2) - s_1 s_2 \sin\delta_1 \sin\delta_2 \\ c_2 s_2 (1 - \cos\delta_2)(c_1^2 + s_1^2 \cos\delta_1) + c_1 s_1 (1 - \cos\delta_1)(c_2^2 + s_2^2 \cos\delta_2) + s_1 c_2 \sin\delta_1 \sin\delta_2 \\ s_2 \sin\delta_2 (c_1^2 + s_1^2 \cos\delta_1) - c_1 s_1 c_2 \sin\delta_2 (c_1^2 + s_1^2 \cos\delta_1)(1 - \cos\delta_1) + s_1 \sin\delta_1 \cos\delta_2 \end{pmatrix} \quad (28)$$

where $c_i = \cos 2\theta_i$, $s_i = \sin 2\theta_i$. To generate the needed four Stokes vectors to be complete, we can play with 10 parameters (the fixed azimuths and the four pairs of retardations); which are far too many ! Actually, among the many possibilities, the theoretical minimum of $c(\mathbf{W})$ is reached for azimuth values given by $\theta_1 = \varepsilon\, 27.4° + q\, 90°$, and $\theta_2 = \varepsilon\, 72.4° + r\, 90°$, where $\varepsilon = \pm 1$ has the same value in both equations, while $q$ and $r$, are any integer numbers (not necessary equal). Retardation sequences the form:

$$(\delta_1, \delta_2) = (\Delta_1, \Delta_1), (\Delta_2, \Delta_1), (\Delta_1, \Delta_2), \text{ and } (\Delta_2, \Delta_2), \quad (29)$$

with optimal values being $\Delta_1 = 315° + p\, 90°$, and , $\Delta_2 = 135° + p\, 90°$ respectively, where again, $p$ is an arbitrary integer.

As retardations can be adjusted on demand, PSGs based on nematic liquid crystals can in principle reach the theoretical minimum of $c(\mathbf{W})$ for any wavelength within their spectral range. This possibility of complete optimization make them very well suited for Mueller ellipsometric measurements discrete wavelengths, provided total acquisition times of the order of 1 sec for the whole set of 16 images is acceptable.

*Ferroelectric Liquid Crystals*
With respect to nematics, ferroelectric liquid crystals (FLCs) feature the following quite different, and complementary, characteristics :
- They are also linear retarders, but with constant retardation. What is driven electrically is the orientation of their fast axis. This *orientation is actually bi-stable*, with two possible azimuths 45° apart from each other. The polarity of the DC driving voltage actually defines which of these two azimuths is actually reached.
- These devices may switch from one state to other *extremely fast*, typically in less than 100 µs.

The commutation speed of these components allow fast Mueller ellipsometry, either in spectroscopic or in imaging modes. However, due to the fixed values of retardations, any PSG built with these components will not allow a fine minimization of the condition number like that possible with nematics. This minimization can be performed only as a compromise over all the spectral range of interest. On the other hand, if acceptable values are obtained throughout this range, with *c* values typically less than 4 or so, then the data can be taken simultaneously over this range, allowing fast spectral ellipsometry and/or "color" Mueller imaging.

We first consider a configuration similar to that described above for nematic LCs. A linear polarizer is followed by two FLCs, which are switched alternatively to actually generate the four needed polarization states. If we now call $\theta_1$ and $\theta_2$ two possible azimuths of the FLCs, when the driving voltages are switched, the resulting pairs of azimuth are

$$\{(\theta_1,\theta_2), (\theta_1+45°,\theta_2), (\theta_1,\theta_2+45°), (\theta_1+45,\theta_2+45°)\} \quad (30)$$

Again, the four generated Stokes vectors can be calculated by putting these azimuths, and the constant retardations $\delta_1$ and $\delta_2$ into Eq. 28. With this configuration, the best result was obtained with $\delta_1 = 90°$ and $\delta_2 = 180°$ (at 510 nm), and $\theta_1 = 70°$, $\theta_2 = 165.5°$. The spectral dependence of the reciprocal condition number $1/c(\mathbf{W})$ obtained with these parameters is shown as the black line on fig. 8. The qualitative criterion defined above, namely $1/c \geq 0.25$ is obeyed in a relatively narrow range, between 450 and 700 nm. This useful spectral range can be significantly extended by adding a true zero order quarter wave plate for 633 nm. With the same values of $\delta_1$ $\delta_2$ (quarter_ and half-wave at 510 nm) the red curve is obtained for $\theta_1 = -10°$ and $\theta_2 = 165.5°$ meaning that the PSG can be used with the same noise propagation as before between 420 nm (limited by the transmission of the FLCs) and 1000 nm.

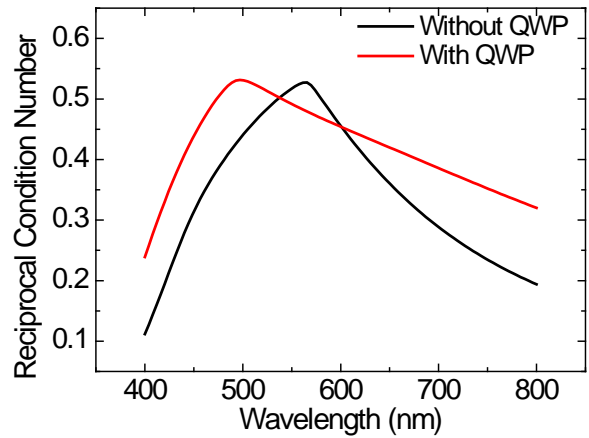

Figure 8. Spectral dependence of the reciprocal condition number $1/c(\mathbf{W})$ of the matrix $\mathbf{W}$ associated to the FLC based PSG. The effect of the insertion of a quartz wave-plate between the FLCs can be clearly seen. Red line with the wave-plate and Black line without it.

**Instrumental Implementations of Mueller Ellipsometers**

**Spectroscopic polarimeters**

Liquid crystal modulation Mueller ellipsometers use liquid crystal devices to modulate the polarization without any mechanical rotations. The first prototype was built in 2003 and presented at the 3$^{rd}$ International Conference of Spectroscopic Ellipsometry (ICSE-III) held in Vienna[34]. The first commercial system appeared in 2005 under the name of MM-16 by HORIBA Scientific. Since then, the product has been further developed, with new versions now available on the market. For instance, a particular implementation adapted to the measurement of small samples, commercialized under the name of AutoSE was launched in 2008. The latest version, called SmartSE, combines spectroscopic and imaging capabilities. The spectral range has been expanded. Initially it was limited to the visible (450 to 850nm), but presently can reach the near infrared (450 to 1000nm). The working spectral range of liquid crystal-based polarimeters is determined by the transparency of the liquid crystal devices. In the short wavelength range, UV radiation must be avoided because it may induce chemical modifications, or even destroy the liquid crystals which are made of delicate organic molecules. In the near infrared, the limitation stems from the thin conducting oxides which are deposited on the windows of the liquid crystal devices to control electrically the orientation of liquid crystals. The conducting oxides have a high concentration of free charge carriers which very efficiently absorb the near infrared (>1500nm) light, making the devices opaque[35].

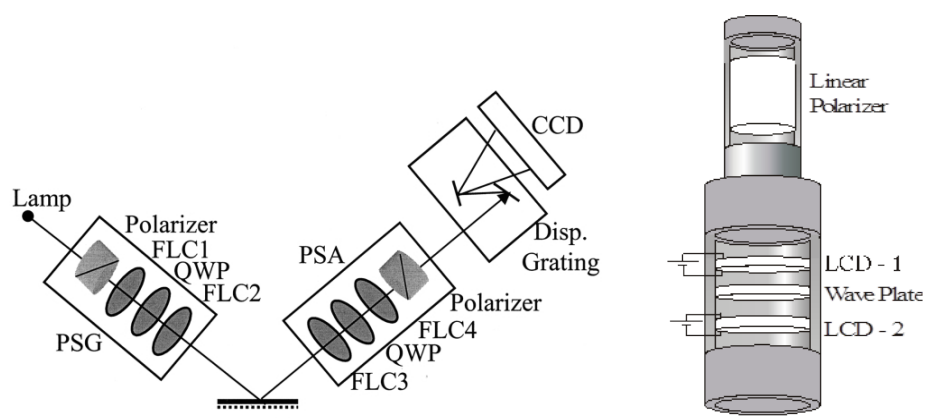

Figure 9. Left: Schematic representation of the general set-up of a Mueller ellipsometer mounted in reflection configuration, showing the PSG, the sample and the PSA. Right: Schematic of the PSG. The PSA is identical to the PSG. Reproduced with permission from Thin Solid Films[32].

The basic configuration of the ferroelectric LCD-based ellipsometer is the PSG, the sample and the PSA, as shown schematically in fig. 9. The resulting configuration of the PSG consists of a linear polarizer, a Glam Thomson, a first ferroelectric LCD device, a true zero order wave plate, and a second ferroelectric LCD device. The quartz wave plate partially compensates for the spectral dependence of the retardation of the LC plates, thus making the condition number as constant as possible along the measured spectral range as shown in fig. 8. The azimuths of the three plates with respect to those of the polarizer are also specified to fit well with our design criterion. The PSA is identical to the PSG, but with its elements in reverse order. As a source of illumination, we use a 30W halogen lamp, and as a detector, we use a CCD array coupled to a commercial HORIBA Scientific diffraction grating optimized to work between 400 and 1000nm. The polarimeter can work in transmission mode as well as in reflection mode. For practical purposes, the PSG and PSA are mounted on an automatic goniometer for variable angles of incidence from 40° to 90° with a step of 0.01°. Thanks to the goniometer, the Mueller ellipsometer can make measurements in reflection mode (<90°) and in transmission mode (90°). The sample holder is mounted on an automated theta table which allows the sample to turn along an axis perpendicular to the surface. This azimuthal movement is interesting for the characterization of anisotropic samples and diffracting structures.

The liquid crystal modulation ellipsometer measures a spectroscopic Mueller matrix in one shot. For isotropic plane surfaces, the Mueller matrix is block-diagonal and their elements are related with the ellipsometric angles Ψ and Δ. Therefore, for this kind of material, the Mueller ellipsometer can be used as a fast standard spectroscopic ellipsometer.

$$M = \begin{pmatrix} 1 & -\cos(2\Psi) & 0 & 0 \\ -\cos(2\Psi) & 1 & 0 & 0 \\ 0 & 0 & \cos(2\Psi)\cos(\Delta) & \cos(2\Psi)\sin(\Delta) \\ 0 & 0 & -\cos(2\Psi)\sin(\Delta) & \cos(2\Psi)\cos(\Delta) \end{pmatrix} = \begin{pmatrix} 1 & -Ic' & 0 & 0 \\ -Ic' & 1 & 0 & 0 \\ 0 & 0 & Ic & Is \\ 0 & 0 & -Is & Ic \end{pmatrix} \quad (9)$$

For anisotropic samples, or diffracting structures oriented at an arbitrary direction with respect to the plane of incidence, the Mueller matrix is no longer block-diagonal but shows a high degree of symmetry.

**Imaging ellipsometers and polarimeters**

This instrument can be seen as the ultimate development of the well-known polarized microscopy, as the polarimetric characterization of the sample is complete, in contrast with the usual setups with crossed linear polarizers or left and right circular polarizers. An overall view of the imaging polarimeter[36,37] is shown in fig. 10.

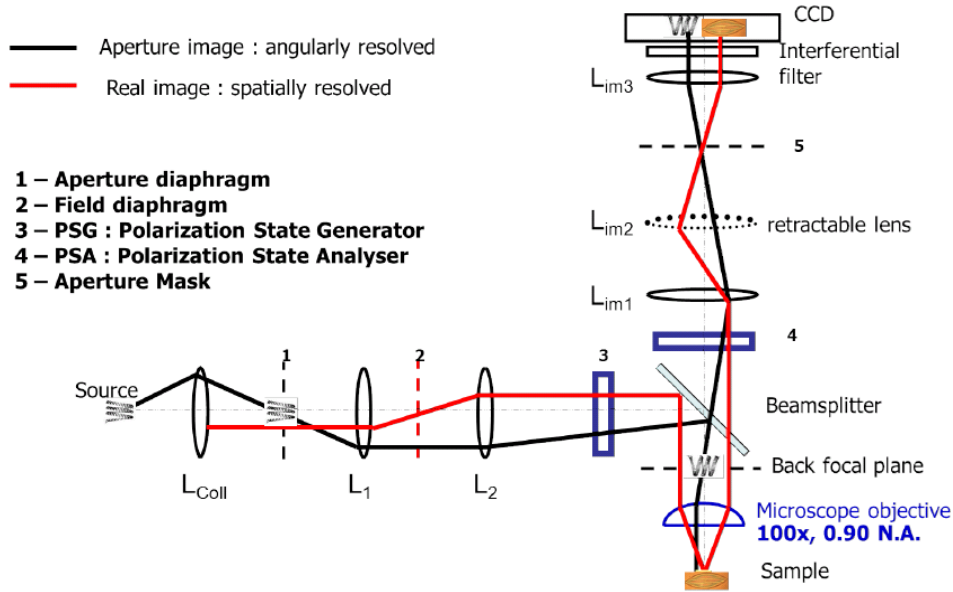

Figure 10. Schematic representation of the imaging/conoscopic Mueller polarimeter.

A microscope objective (Nikon Plan Achromat 100×) with a high numerical aperture (0.90) is illuminated by a halogen source via a fiber bundle followed by an input arm comprising, among other elements:
- an *aperture diaphragm*, which is imaged on the objective back focal plane (BFP), and is used to define the *angular distribution* of the light incident on the sample,
- a *field diaphragm*, imaged on the sample, which allows to define *the size of the illuminated area* on the sample.
- The PSG, to modulate the incident polarization,

- A *nonpolarizing beamsplitter*, with approximately 50% transmission and reflection coefficients, to steer the beam onto the microscope .

On the detection side, we find
- The *beamsplitter* again, working this time in transmission,
- The PSA to analyse the emerging polarization,
- A set of two lenses which *image the objective back focal plane onto a two dimensional* imaging detector,
- A "retractable" lens which can be inserted in the beam path t*o image the sample* instead of the objective back focal plane,
- An *aperture mask* can be set in a plane conjugated with the objective BFP, typically to eliminate some strong contributions in order to see weaker ones, or to select the visualized diffraction orders if the sample is a grating.
- An interferential filter, typically quite narrow for metrological applications.
- The camera, a backthinned and cooled 512x512 pixel CCD from Hamamatsu

The PSG and PSA operating in this setup used nematic liquid crystals because the samples were static, in consequence no need to the fast commutation times of the ferroelectric crystals, and we considered it was important to be able to minimize the condition numbers at each wavelength[34, 38].

The angular distribution and the spot size characterizing the beam incident onto the sample can be adjusted independently of each other (at least as long as the illumination beam is far from being diffraction limited, a condition which is always fulfilled in practice with the classical light sources such as the one we use). The two modes of operation of the microscope are illustrated on fig. 11. The left panel shows the real space image of image of a grating, and the reciprocal space image obtained with a slit as an aperture diaphragm and the grating as the sample : due to the presence of the slit, the angular distribution of the incident light is almost 1D, which is then diffracted in orders 0 (central line) and +-1 (lateral lines).

The right panel of fig. 11 shows how the angular distribution of the light coming from the sample is actually mapped on the objective back focal plane. Due to Abbe's sine condition[5] a parallel beam emerging from the sample with a polar angle $\theta$ and an azimuth $\phi$ is focused in the back focal plane on a point with radial coordinates ($f \sin\theta, \phi$), where f is the objective focal length. Of course, while all the azimuths between 0° and 360° are mapped, the polar angles $\theta$ are limited by the numerical aperture of the objective. In our case the nominal values are $\sin \theta_{max} = 0.90$ and $\theta_{max} = 65°$. In practice, it is difficult to achieve the full angular range. The radial coordinate is calibrated on the images by using diffraction patterns obtained with gratings with known pitches, such as the pattern shown in the left panel of fig. 11, and our images are limited to about 60°.

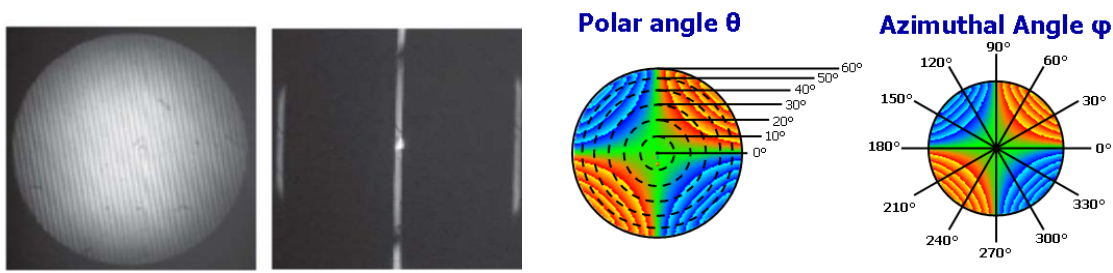

Figure 11. *Left*. Real-Fourier space images of a grating. *Right* Angular coordinates in the Fourier Space (maximum aperture 62°). Images reproduced with permission from Physica Status Solidi A[34].

Imaging in reciprocal space may constitute an interesting alternative to the more conventional approach of goniometric ellipsometry/polarimetry if angularly resolved data are to be acquired. Measurements along the polar angle at a fixed azimuthal angle of an image are equivalent to measurements taken at different angles of incidence on a non-imaging system. Accordingly, measurements recorded at a fixed polar angle and along the azimuthal direction on an image are equivalent to measurements taken rotating the sample holder in a non-imaging ellipsometer. With respect to simple conoscopy through crossed polarizers, full polarimetric conoscopy can be very useful to characterize anisotropic materials, as it provides angularly resolved maps of retardation (and diattenuation, if present) which significantly constrains the values of the dielectric tensor from easy and fast measurements[39]. Moreover, under a powerful microscope objective, the *spot size can easily be reduced to* 10 *µm or less*, a possibility which can be very useful for some metrological applications, and more particularly in microelectronics. Obviously, so small spots sizes are much more difficult to obtain with the usual ellipsometric setups involving narrow beams with low numerical apertures.

As an example of Mueller images in the Fourier space we show in left panel of fig. 12 the data taken on a silica thick plate. At first sight the observed patterns may seem surprising for an isotropic sample. In fact, the isotropy is "broken" by the choice of the basis used to define the polarization, and which is uniform all over the image with one vector horizontal and the other vertical. Obviously, these are not the usual (*p,s*) vectors defined with respect of the incidence plane, and which would be oriented radially in each point of the image.

In the same figure, there are shown the angularly resolved values of Δ and Ψ deduced from the experimental Mueller matrix (top), together with the corresponding simulations (bottom). As expected, both parameters display an almost perfect radial symmetry. Moreover Δ remains at zero, and then "jumps" to 180° at the Brewster incidence, while Ψ starts at 45° at the image center, and then decreases in agreement with the theory (the jump from red to light yellow indicates that the plotted value went below the minimum of the scale, here 15°).

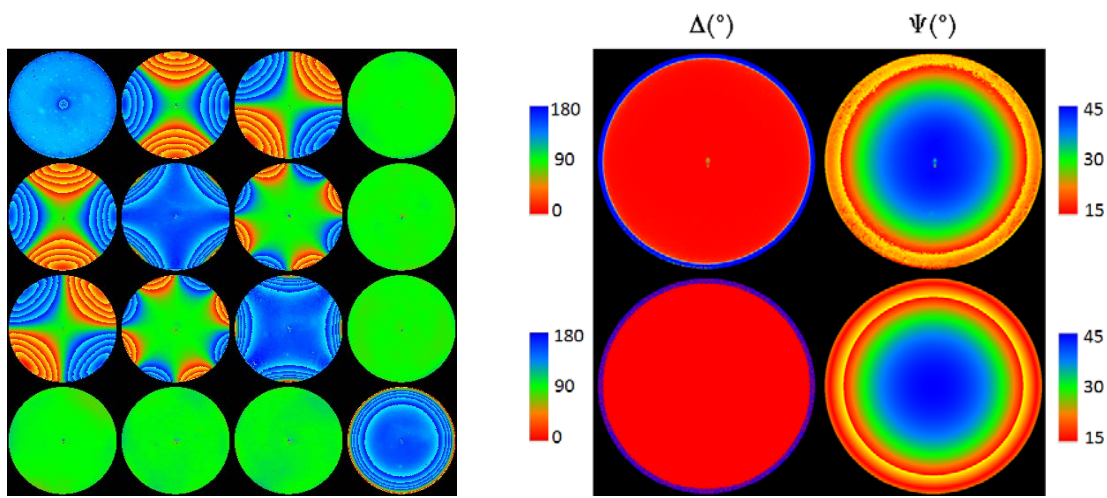

Figure 12. *Left* : raw Mueller images in the reciprocal space of a thick plate of silica. The basis vectors for the definition of polarisaton are vertical and horizontal all over the image.

*Right*:, *top*: maps of angularly resolved Δ and Ψ derived from associated to the data shown in the left panel. *Right bottom* :corresponding simulations.

These results clearly show that this technique may be very powerful. In the following section we give an example of application for the metrology of sub-wavelength gratings. However, it would be extremely difficult to "push" the accuracy of such measurements to the levels reached by usual, non-imaging ellipsometers. The main reason for this is that the objectives used in Fourier configuration may introduce some polarimetric artifacts which cannot be taken into account by the ECM method, as the system must be calibrated with the objective removed[36]. Moreover, even strain free objectives are extremely sensitive to mechanical constraints, and the resulting artifacts would probably evolve in time. In spite of these limitations in accuracy, Mueller microscopes (operating here in reflection, but transmission may be used too) are likely to open new research topics in many areas.

**Examples of Mueller Ellipsometric Measurements**

The benefit of using Mueller ellipsometers lies in the characterization of samples with a complex optical response; e.g., anisotropic layers or diffracting structures. In this section we focus on the profile reconstruction and overlay characterization of diffraction gratings. Optical methods (also called "scatterometry") are fast, non-destructive, and may exhibit strong sensitivity to tiny changes in grating profiles[40]. As a result, they are becoming increasingly popular for process control in the microelectronics industry[41]. On the other hand, these methods are indirect, and the reconstructed profiles may depend on the model used to fit the data (and the dielectric function of somewhat "ill-defined" materials like resists.) Possible model inadequacies do not necessarily appear in the goodness of fit. Parameter correlations may also constitute a serious issue, as shown in a comprehensive study of the results of scatterometric reconstruction by the usual techniques (normal incidence reflectometry and planar diffraction spectroscopic ellipsometry) of various profiles representing different technological steps[42]. In this context, Mueller ellipsometry may constitute an interesting alternative, provided the data are taken in conical diffraction geometries. In conical diffraction geometries, the symmetry axes of the grating structure are neither parallel nor perpendicular to the plane of incidence. Indeed, in such geometries, the grating Jones matrix is no longer diagonal (and the Mueller matrix no longer block-diagonal). As a result, additional information is available and may help in constraining the fitting parameters. Moreover, the stability of the optimal values of these parameters when the azimuth is varied may constitute a much better test of the model relevance than goodness of fit at a single azimuth[43]. Angle resolved scatterometry, with a high numerical aperture microscope objective, as described above[37,44], may also constitute an interesting scatterometric tool, as it greatly facilitates measurements in extremely tiny targets (less than 5 μm wide), an increasing requirement by semiconductor manufacturers. This would be particularly true for *overlay* (default of positioning of superimposed grids at different layers), a parameter which is becoming increasingly critical and will require in-die dense sampling while current methods involve up to 8 standard (50 μm wide) targets to provide all the relevant information[45]. In the following we provide two examples. The first illustrates the use of spectroscopic measurements, and the second shows the possibilities of the angle-resolved polarimeter.

**Profile reconstruction using spectroscopic Mueller ellipsometry**

In the following we will show the results of a study whose goal was to show the possibility of using Mueller ellipsometry data for reconstruction (optical metrology) of a diffraction gratings profile. The sample analyzed consisted of a silicon wafer with a series of silicon gratings etched on it using UV beam lithography. Typical dimensions for the gratings were: groove depths around 100nm, line widths around 130nm and 250nm, and pitches from 500 to 1100nm. Each individual grating was etched in an area of 3x3mm. Etched silicon gratings were chosen for this study because of their long term dimensional stability, higher refractive index contrast and relevance to semiconductor industry. For reference, the dimensions of the profiles of the gratings were determined by a state-of-the-art 3D AFM microscope. For the sake of simplicity, we show the results corresponding to only one grating. For more details, please refer to[43,47]. Experimental data were taken by a HORIBA Scientific MM-16 Mueller Ellipsometer, operating in the visible (450 – 850nm)[34]. A series of measurements were taken varying the azimuth over 360° in steps of 5°. The incident angle was kept at 45° to make sure the beam diameter on the sample was small enough to safely maintain the spot within the grating. Two measured spectroscopic matrices, corresponding to azimuthal angles +45° and -45°, together with the corresponding fits, are shown in fig. 13. The matrix elements are normalized by the element $M_{1,1}$, and thus vary from -1 to 1. The redundant information in the Mueller matrix allows us to evaluate the quality of the measured data by simple criteria such as the degree of polarization in Eq. 7 or the symmetry of the off-diagonal elements. In the upper left corner of fig. 13 is a plot of the degree of polarization, which was found to be very close to 1, an indication of zero depolarization due to the high quality of both sample and data. The blue and green spectra coincide in the diagonal blocks, while they are opposite in the off-diagonal blocks. These symmetries provide a robust test of the accuracy of both the measurements and the sample azimuthal position.

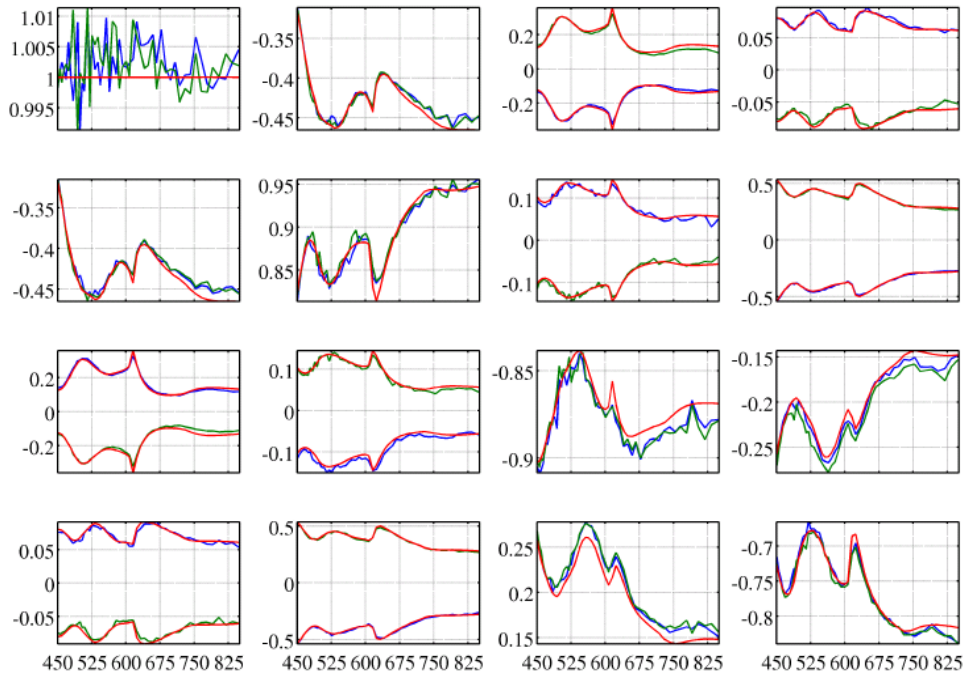

Figure 13. Measured (blue and green lines) and fitted data (red and orange lines) at two different azimuthal angles of ±45° and incidence angle of 45°. Figures reproduced with permission from SPIE Proceedings[44].

The measured data were fitted by RCWA simulations[46] formulated in the Mueller–Jones formalism[47]. The profile of the gratings was represented using a different models. For the sake of clarity here we cite only two models. The first model assumed the profile to be trapezoidal. The second model the grating profile was represented by the superposition of two rectangular lamellas. Both models are sketched in fig 14. The trapezoidal model depends on three adjustable parameters, the thickness, $d$, the CD (with) of the lines, and the trapeze angle (SWA). The second model depends on four parameters, the CDs and the thickness of the two lamellas. The resulting best-fitted parameters for both models are presented in fig. 14. In general both models provided fits of same quality, but the most prominent difference among them was the dependence of the best-fitted parameters with the azimuth angle at which the measurements were taken. Whereas best-fitted parameters corresponding to the model of two lamellas showed a low dependency with the azimuthal angle, CD2 values are dispersed by less than 1.5 nm and the grating depth varies by less than 2 nm around 108 nm, the parameters fitted with the trapezoidal model showed strong fluctuations, 5 nm for the CD and 10 nm for the thickness. The second element that makes the difference between both models is the correlation between fitted parameters. A close look to the values of the CD and the thickness corresponding to the trapezoidal model reveals that them are strongly linearly correlated. This means that the data does not carry the information needed by the model to discriminate the particular influence of each parameter. In contrast, regarding the bi-lamellar model only a small correlation between the overall grating depth and the bottom lamella depth can be observed in this figure. The low amount of correlation and dependency of the fitted parameters with the observation conditions, show that the bi-lamellar model represented better the profile than the trapezoidal model. The adequacy of the model was also confirmed by comparing the obtained profile with AFM measurements. Similar results have been obtained on all the gratings of the sample.

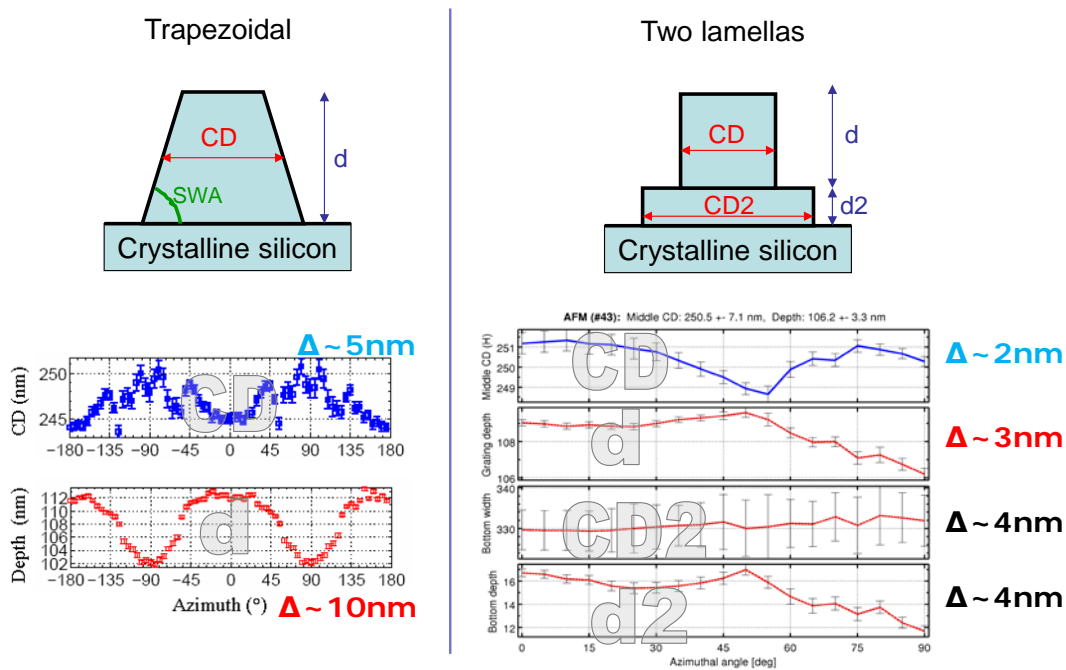

Figure 14. *Left top*. Profile representing the trapezoidal model with the two characteristic parameters, the CD, the thickness and the SWA. *Left bottom*. Resulting best-fitted values for the parameters CD and thickness at different azimuths. The corresponding variations are indicated by the symbols Δ. *Right top*. Profile representing the double lamella model with the four characteristic parameters. *Right bottom*. Results of the fit of four free parameters of two lamellas model over different azimuthal angles. Error bars in figures denote statistical errors. The maximum variation of each parameter is indicated with the symbol Δ. Images reproduced with permission from SPIE Proceedings[43].

In summary, this example shows that spectroscopic Mueller Ellipsometry is a non-destructive and accurate technique for studying grating profiles. Mueller Ellipsometry has the advantage of being faster and cheaper than other tests currently used for in-line quality control in the microelectronics industry.

**Overlay characterization using angle resolved Mueller imaging ellipsometry**
The overlay is defined as the misalignment between two layers of a stack. This error could lead to defective transistors, for example, if there is no electrical contact between the different constitutive layers. This feature used to be of no interest because its effect was negligible when compared to the defects in critical dimension. With the shrinking of the technology node (TN), overlay control is becoming more and more critical in semiconductor manufacturing. If this overlay is higher than a set threshold, the whole batch cannot be processed to the new step. This results in a rework, meaning the wafer is returned to the previous lithography step and the resist is stripped. In the case of grating profile optical metrology, there are several techniques that are considered as a reference for the microelectronic industry. Those techniques include non-optical techniques, such AFM or SEM microscopy, and optical techniques based on image analysis (pattern recognition) and on scatterometry. Image analysis, also known as Advanced Image Metrology (AIM), is used in this work as reference to check the quality of the results obtained by angle- resolved Mueller ellipsometry.

The choice of proper azimuthal configuration for the measurements with spectroscopic polarimetry is extremely important for the overlay characterization[48-49]. Given that the angle resolved polarimeter gives an angular signature, it is possible to use the symmetries of the grating to enhance the sensitivity of its angle-resolved signature. The sign of the off-diagonal blocks of the measured Mueller matrix changes when the azimuth $\varphi$ is changed into -$\varphi$. If the profile is symmetric, the signature is invariant when $\varphi \rightarrow \varphi+180°$ and also for the special case of $\varphi = 90°$, the previous two conditions can only be fulfilled if the off-diagonal blocks are zero. A rupture of symmetry in the structure will violate the above conditions and the off-diagonal blocks will take non-zero values for $\varphi = 90°$. Moreover, given that these blocks change sign upon a mirror symmetry, the information about the sign of the overlay can be unambiguously extracted. In order to highlight the influence of the overlay over the off-diagonal elements of the Mueller matrices, the following estimator was defined: $\mathbf{E} = |\mathbf{M}|-|\mathbf{M}|^T$ where the superscript T denotes the transposed matrix. The estimator works well with either 1D or 2D gratings, and for different types of overlays.

For the sake of clarity, below is a simple example. It consists of the overlay of a 1D grating. As depicted in fig. 15, the overlay is the small shift defined along the direction perpendicular to the lines of the grating. For this particular example with an overlay of 25nm, the elements of the estimator matrix $\mathbf{E}$ can reach the value of 0.25 (m14 and m41), i.e. 1/8 of the total scale, which points out the high sensitivity of this estimator.

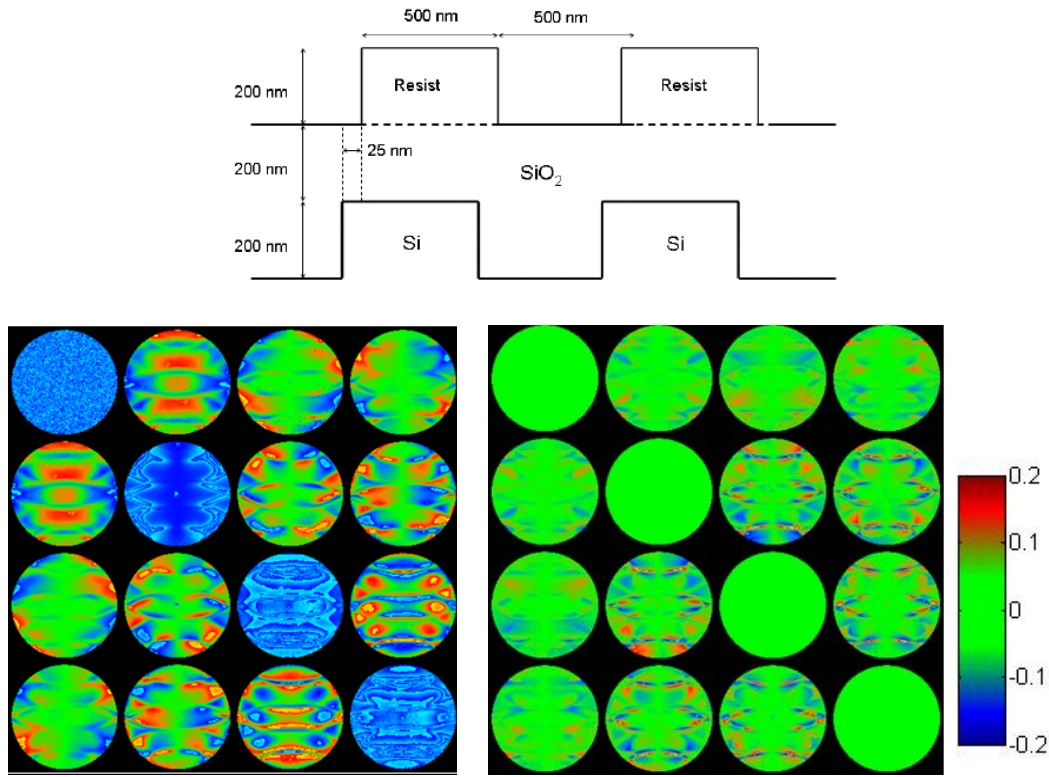

Figure 15. Schematic view of the grating used in the experiences for the overlay characterization. The overlay: 25nm. *Bottom-Left*. Experimental angle-resolved Mueller matrix. *Bottom-Center*. Corresponding estimator matrix **E**. *Bottom-Right*. Color Scale. Images reproduced with permission of the author[49].

To check the linear relation between the values of the estimator **E** and the value of the overlay, we compared the maximum value of the element $E_{1,4}$ of the estimator matrix with the overlay value obtained by AIM for a set of samples. The results are shown in fig. 16.

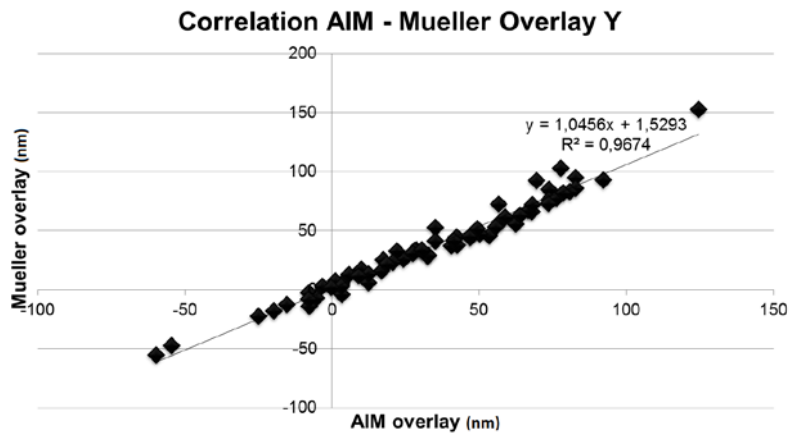

Figure 16. Correlation between measured and AIM Y-overlay for 55 samples with different overlays. Figure reproduced with permission of the author[50].

The fitted linear regression, also shown in fig. 16, is excellent. Moreover, the figure indicates that these results can be extended to negative overlays, with the sign of the estimator becoming negative.

In conclusion, it can be said that the overlay errors can be accurately determined, provided that we are able to design an apparatus with small enough errors. For this particular case, the magnitude of the systematic errors of the experimental matrices was estimated to be on the order of 1%, which was associated to an incertitude of about 1nm in the determination of the overlay.

**Conclusions**

This article focuses on the state-of-the-art standard ellipsometric and new Mueller ellipsometric techniques, as well as their applications. The theoretical approaches necessary to properly describe these techniques, including the most general Stokes-Mueller formalism, were briefly described. A series of ellipsometric techniques were then reviewed, including spectroscopic and imaging Mueller setups. We have shown several examples to illustrate the high sensitivity of Standard and Mueller Ellipsometry and their adequacy for practical applications, such as material characterization, or dimensional metrology for process control in material science, microelectronics, and the solar industry, to name a few. Indeed, for many applications (process control, biomedicine, etc.) the intrinsically fast and non-invasive ellipsometric techniques may be implemented at much lower costs than other "reference" techniques (such as TEM, AFM, and SEM imaging for nanostructures, for example). As a result, either standard or Mueller ellipsometric techniques offer great potential for significant development in many economically important activities.

**FIGURE CAPTIONS**

Figure 1. Examples of electric field trajectories in the plane perpendicular to the propagation direction for fully polarized (*left*) or partially polarized (*right*) light waves.

Figure 2. Schematic view of an ellipsometric measurement in reflection configuration. The polarized beam is incident on the sample from the right. After reflection, the polarization state of the beam is changed and light propagates to the left.

Figure 3. Schematic representation summarizing the different parameters related to the sample that can be deduced using ellipsometry. These parameters include: thin film thickness, optical constants, roughness, porosity, composition, uniformity, etc.

Figure 4. General scheme of a standard ellipsometer. The PSA is the Polarization State Analyzer, which distinguishes the various optical configurations described in this section.

Figure 5. Simulated Ψ (blue) and Δ (red) angles over the spectral range from 200 to 1700nm, for a c-Si substrate covered with a thin layer of $SiO_2$. The different spectra (clearly seen in Δ) correspond to different thickness of the $SiO_2$ layer.

Figure 6. Simulated Ψ (blue) and Δ (red) values over the spectral range from 200 to 1700 nm, for a glass substrate covered with a thin film of $SiO_2$. The different spectra correspond to different thicknesses of the $SiO_2$ layer varying from 0 to 10 nm by a step of 1 nm.

Figure 7. General principle of operation of any Mueller Ellipsometry.

Figure 8. Spectral dependence of the reciprocal condition number $1/c(\mathbf{W})$ of the matrix $\mathbf{W}$ associated to the FLC based PSG. The effect of the insertion of a quartz wave-plate between the FLCs can be clearly seen. Red line with the wave-plate and Black line without it.

Figure 9. *Left*: Schematic representation of the general set-up of a Mueller ellipsometer mounted in reflection configuration, showing the PSG, the sample and the PSA. *Right*: Schematic of the PSG. The PSA is identical to the PSG. Reproduced with permission from Thin Solid Films[32].

Figure 10. Schematic representation of the imaging/conoscopic Mueller polarimeter.

Figure 11. *Left*. Real-Fourier space images of a grating. *Right* Angular coordinates in the Fourier Space (maximum aperture 62°). Images reproduced with permission from Physica Status Solidi A[34].

Figure 12. *Left* : raw Mueller images in the reciprocal space of a thick plate of silica. The basis vectors for the definition of polarisaton are vertical and horizontal all over the image. *Right:, top*: maps of angularly resolved Δ and Ψ derived from associated to the data shown in the left panel. *Right bottom* :corresponding simulations.

Figure 13. Measured (blue and green lines) and fitted data (red and orange lines) at two different azimuthal angles of ±45° and incidence angle of 45°. Figures reproduced with permission from SPIE Proceedings[44].

Figure 14. *Left top*. Profile representing the trapezoidal model with the two characteristic parameters, the CD, the thickness and the SWA. *Left bottom*. Resulting best-fitted values for the parameters CD and thickness at different azimuths. The corresponding variations are indicated by the symbols Δ. *Right top*. Profile representing the double lamella model with the four characteristic parameters. *Right bottom*. Results of the fit of four free parameters of two lamellas model over different azimuthal angles. Error bars in figures denote statistical errors. The maximum variation of each parameter is indicated with the symbol Δ. Images reproduced with permission from SPIE Proceedings[44].

Figure 15. Schematic view of the grating used in the experiences for the overlay characterization. The overlay: 25nm. *Bottom-Left*. Experimental angle-resolved Mueller matrix. *Bottom-Center*. Corresponding estimator matrix $\mathbf{E}$. *Bottom-Right*. Color Scale. Images reproduced with permission of the author[49].

Figure 16. Correlation between measured and AIM Y-overlay for 55 samples with different overlays. Figure reproduced with permission of the author[49].

# Tables

Table I. Summary of essential characteristics, advantages and weaknesses of the main ellipsometric techniques.

| Technology | Modulation (Hz) | Measured Parameters | Strengths | Weakness |
|---|---|---|---|---|
| **Null** | 0 | $\Psi$ and $\Delta$<br>9 Mueller matrix elements | • Simple optical assemblies<br>• High accuracy and precision relatively easy to achieve | • Spectroscopic extension difficult<br>• Sensitive to residual polarization from source or detector |
| **Rotating Polarizer Analyser** | Several to hundreds | $\tan(\Psi)$ and $\cos(\Delta)$<br>9 Mueller matrix elements | • Simple optical assemblies<br>• Quasi-achromatic instruments<br>• Wide spectral range of operation<br>• Easy CCD detection | • Inaccurate measurements for $\Delta$ approaching 0° or 180°<br>• Does not measure the *V* component of the Stokes vector.<br>• Sensitive to residual polarization from source or detector |
| **Rotating Compensator** | Several to hundreds | $S_2 = -\cos(2\Psi)$<br>$S_3 = \sin(2\Psi)\cos(\Delta)$<br>$S_4 = \sin(2\Psi)\sin(\Delta)$<br>12 Mueller matrix elements | • Complete measurement of the Stokes vector from a single detection configuration<br>• EasyCCD detection<br>• Generalized ellipsometry possible Depolarization + 12 elements of the Mueller matrix | • Complicated optical assemblies<br>• Rotating may introduces inaccuracies because of beam wandering due to compensator imperfections<br>• Complex calibration procedures |
| **Phase-Modulation** | $50 \cdot 10^3$ to $100 \cdot 10^3$ | $S_2 = \sin(2\Psi)\cos(\Delta)$ ; conf. (II)<br>$S_3 = \sin(2\Psi)\sin(\Delta)$ ; confs (II) & (III)<br>$S_4 = \cos(2\Psi)$ ; conf. (III)<br>12 Mueller matrix elements | • No rotating elements & Fast measurements.<br>• Excellent signal-to-noise ratio from VUV to NIR<br>• Accurate measurement of ($\Psi,\Delta$) with two detection configurations (II) & (III)<br>• Generalized ellipsometry possible Depolarization + 12 elements of the Mueller matrix, | • Temperature sensitive photoelastic modulator<br>• No "CCD" detection<br>• Chromatic dependence of the photoelastic-modulator<br>• Measurements of the 12 Mueller matrix elements require multiple (6 to 8) runs |

Table II. The different thickness values together with the calculated $\Psi$ and $\Delta$ angles at 633nm wavelength.

| $\Delta$ (°) | $\Psi$ (°) | Film thickness (Å) |
|---|---|---|
| 179.195 | 10.567 | 0 |
| 178.897 | 10.568 | 1 |
| 178.599 | 10.568 | 2 |
| 178.302 | 10.569 | 3 |
| 178.004 | 10.571 | 4 |
| 177.706 | 10.572 | 5 |
| 177.409 | 10.573 | 6 |
| 177.111 | 10.575 | 7 |
| 176.814 | 10.577 | 8 |
| 176.516 | 10.579 | 9 |
| 176.219 | 10.582 | 10 |

Table III. Calculated Ψ and Δ vales at 633nm and 190 nm for 0 and 10 nm SiO$_2$ layer. The table also includes an estimation of the absolute experimental uncertainty of Ψ and Δ at 633nm and 190nm. The ratio between the differences among the Ψ and Δ measured at a thickness of 0 and 100 nm respect to the uncertainty in the measurements, give an idea of the sensitivity of Ψ and Δ at different wavelengths.

| Thickness (nm) | λ 190 nm | | λ 633 nm | |
|---|---|---|---|---|
| | Ψ | Δ | Ψ | Δ |
| 0 | 17.78 | 0.038 | 20.34 | 0.001 |
| 10 | 18.20 | 3.233 | 20.37 | 0.861 |
| Diff. | 0.422 | 3.195 | 0.03 | 0.86 |
| Absolute error | 0.01 | 0.01 | 0.005 | 0.005 |
| Diff / Error error | 42 | 320 | 6 | 170 |